\documentclass[12pt,preprint]{aastex}
\begin{document}

\title{Semi-Analytical Models for Lensing by Dark Halos:\\ I. Splitting 
Angles}

\author{Li-Xin Li and Jeremiah P. Ostriker}
\affil{Princeton University Observatory, Princeton, NJ 08544--1001, USA}
\email{E-mail: lxl, jpo@astro.princeton.edu}

\begin{abstract}
We use the semi-analytical approach to analyze gravitational lensing of 
remote quasars by foreground dark halos in various cold dark matter (CDM)
cosmologies, in order to determine the sensitivity of the predictions for 
probabilities of images separations to the input assumptions regarding 
the properties of halos and cosmological models. The power spectrum of 
primordial fluctuations is normalized by the cluster abundance 
constraints. The mass function of dark halos is assumed to be given by 
the Press-Schechter function. The mass density profile of dark halos is 
alternatively taken to be the singular isothermal sphere (SIS), the 
Navarro-Frenk-White (NFW) profile, or the generalized NFW profile. The 
cosmologies being considered include: the Einstein-de Sitter model (SCDM), 
the open model (OCDM), and the flat $\Lambda$-model (LCDM). As expected, 
we find that the lensing probability is extremely sensitive to the mass 
density profile of lenses (dark halos), and somewhat less so to the mean 
mass density in the universe, and the amplitude of primordial fluctuations.
NFW halos are very much less effective in producing multiple images than 
SIS halos. For NFW lenses, the SCDM model produces lensing events fewer 
than the OCDM/LCDM models by two orders of magnitude. For 
SIS lenses, the SCDM model produces more lensing events with small 
splitting angles but produces fewer lensing events with large splitting 
angles than the OCDM/LCDM models, which is due to the fact that for large 
mass halos the Press-Schechter function is very sensitive to the amplitude 
of primordial fluctuations. In all cases the difference between the OCDM 
model and the LCDM model is not dramatic. None of these models can 
completely explain the current observations: the SIS models predict too 
many large splitting lenses, while the NFW models predict too few small 
splitting lenses. Essentially, the observed high ratio of small splitting
to large splitting lenses is not predicted correctly.
This indicates that there must be at least two populations of halos in the 
universe: small mass halos with a steep inner density slope and large mass
halos with a shallow inner density slope. A combination of SIS and NFW 
halos can reasonably reproduce the current observations, if we choose the 
mass for the transition from SIS to NFW to be $\sim 10^{13}$ solar masses
as might plausibly occur due to baryonic cooling and contraction in lower
mass systems. Additionally, there is a tendency for CDM models to have too 
much power on small scales, i.e. too much mass concentration. From our 
sensitivity studies it appears that the cures proposed for other apparent 
difficulties of CDM would help here as well, an example being the warm 
dark matter (WDM) variant which is shown to produce large splitting lenses 
fewer than the corresponding CDM model by one order of magnitude.
\end{abstract}

\keywords{cosmology: gravitational lensing --- galaxies: clusters: general
--- galaxies: halos}

\section{Introduction}
Gravitational lensing directly probes the mass distribution in the universe, 
so the investigation of lensing events of quasars at high redshifts can 
provide us with important information about cosmology and the matter 
distribution in the universe. A number of groups have attempted to test 
cosmological models by comparing lensing probabilities predicted by various 
cosmological models with those obtained from observations 
\citep{tur84,nar88,cen94,wam95,mao97,wam98,bar98,mor00,por00,kee01}. In the 
statistical study of cosmological gravitational lensing, two different 
approaches are used: one is to study the lensing probability of splitting 
angles of multiple images 
\citep{tur84,nar88,cen94,koc95,wam95,wam98,por00,kee01}, 
the other is to study the lensing probability of the length-to-width ratio 
of arcs formed by gravitational lensing \citep{bar94,bar98,coo99,kau00,men00}.
The two approaches reflect two different aspects of gravitational lensing
and are complimentary to each other, both of them are directly related to
observations and deserve detailed investigations. In this paper we focus on 
the splitting angles of multiple images caused by gravitational lensing. 
The case of length-to-width ratio of arcs will be considered in another 
paper. Results of the work done in testing cosmological scenarios to date 
are inconclusive with no clear preference found for any of the current
models, and, overall, one could conclude that none of the studied models
provides a good fit to the rapidly improving observations. Perhaps the 
reason for this indeterminacy is that lensing also 
probes small scale structure. While this is known, its importance has 
perhaps not been appreciated sufficiently. Gravitational lensing provides 
an exquisitely sensitive test to the high frequency part of the power 
spectrum, because it is this part which establishes the central profiles of 
dark matter structures. Since lensing is comparably sensitive to input 
assumptions concerning the properties of lenses and of cosmological models, 
it is hazardous to use observed lensing statistics to draw inferences with 
regard to cosmology before determining the sensitivity to other factors. 
The primary purpose of this paper is to quantitatively assess the 
sensitivity of lensing expectations to the various input parameters so that 
this tool can be used to greatest effect.

To determine the probability for gravitational lensing, we need to know the
mass density profile of lenses (dark halos), the mass function of lenses 
(which gives the number distribution of lenses over masses),
the structure and global features of the universe (the universe is flat,
closed, or open; the matter contents in the universe, etc), and the
positions and shapes of the source objects (quasars). For studying the 
splitting angles of multiple images, it is enough to assume the source 
objects are points at high redshifts. In the framework of the 
Friedmann-Robertson-Walker models, there are a number of cosmologies as
candidates for describing our universe, among which the most popular ones 
are the Einstein-de Sitter universe, which assumes a zero curvature and a 
zero cosmological constant; the open universe, which assumes a negative 
curvature and a zero cosmological constant; and the flat $\Lambda$-universe, 
which assumes a zero curvature and a positive cosmological constant
\citep{pee93,ost95}. The matter contents in the universe are usually 
assumed to be dominated by dark matter: cold dark matter, warm or hot dark 
matter, or the mixture of them \citep{kol90,pee93}. Different kinds of 
dark matter give rise to different mass density profiles for dark halos. If
the matter contents in the universe are dominated by cold dark matter
(CDM), the three cosmological models are denoted respectively as SCDM
(standard CDM = Einstein-de Sitter + CDM), OCDM (open + CDM), and LCDM
(cosmological constant $\Lambda$ + CDM).

The primordial fluctuations determine the mass distribution and the number 
density of dark halos in the universe. In studying gravitational lensing, 
the cosmological models, the matter contents in them, the power spectrum 
of the primordial fluctuations, and the source object positions are 
pre-assumed. Then the mass density fluctuations evolve according to the 
Einstein equations, which can be traced with numerical simulations. As time 
goes on, gravitationally bound objects (dark halos) are formed, which 
play the role of lenses. The mass density profile, and the mass function of 
dark halos in the universe are automatically derived from numerical 
simulations. Then, tracing the trajectories of light rays, we can obtain 
the number of lensing events and thus the lensing probability. This is the 
basic 
spirit of numerical simulations for gravitational lensing \citep{wam95}. 
With this approach, the mass density profile of lenses and the mass function 
of lenses appear as the results of simulations, they do not need to be
pre-assumed. However, numerical simulations have their own limitations. 
First, numerical simulations usually take a great deal of computer time,
which makes it inconvenient to test several different models at a time. 
Second, in every numerical simulation there are two limits on the spatial 
scales: the size of box, which is the maximum scale and determines the 
presence or absence of intrinsically rare events, and the size of 
fundamental cell (i.e. the resolution), which is the minimum scale. 
Correspondingly this gives rise the upper and lower bonds on the masses of
halos formed in the simulation, which causes the result that the number of
lensing events with small splitting angles (caused by halos with small 
masses) and  large splitting angles (caused by rare halos with large 
masses) are underestimated. This last deficit is seen by comparing the
simulation results of \citet{wam95} with the analytical results of 
\citet{koc95}, which will also be confirmed by the results in this paper. 
Furthermore, the lensing cross-sections are very sensitive to inner most
parts of the density profiles which are most subject to problems of 
numerical resolution \citep{ghi00,kly00}. Ultimately, numerical simulations
are required to best determine the expected halo properties. The current
generation of numerical simulations has insufficient dynamical range to 
simultaneously address accurately both large and small scale structures, but
forthcoming $1024^3$-particle simulations may remedy this defect. For the 
present, semi-analytical methods may be best adapted to explore the 
sensitivity of the results to the input parameters.

In the semi-analytical approach, the mass density profile and the mass 
function of lenses are pre-assumed and can be adjusted to fit
various dark matter scenarios. Though this makes us rely on an assumed 
density profile and mass function, it has the advantage that the 
calculation of the lensing probability can be handled with semi-analytical
methods, which takes much less computer time than numerical simulations, so
several different models can easily be tested at a time. And, in this 
approach, there are not the limitations in spatial scales and mass scales
necessitated by numerical simulations; the lensing probability for small
splitting angles and large splitting angles can be obtained accurately for
any assumed halo density profile, the most critical feature of which is the
steepness of the central density profile which may be parameterized by the
central slope, $-\alpha$. The mass density profile is often taken to be the 
singular isothermal sphere (SIS, $\alpha = 2$) \citep{got74}; the mass 
function is usually taken to be the Press-Schechter function \citep{pre74}. 
Though the SIS profile is simple, it does not fit the CDM simulation 
results well. \citet{nav96,nav97} have proposed a ``universal'' density 
profile (the so-called NFW profile, $\alpha=1$) which 
has been shown to fit the simulation results better. A generalized NFW
(GNFW) profile with somewhat steeper inner slopes intermediate between NFW 
and SIS appears to provide a still better fit to simulated profiles
\citep{zha96,sub00,ghi00,jin00,wyi00} for which typically $\alpha = 1.4
\pm 0.1$. With N-body simulations of the SCDM scenario \citep{efs90,kau93}, 
or comparison with observations \citep{bah93,gir98,gir00}, the 
Press-Schechter mass function has been shown to describe the mass 
distribution in the universe quite accurately.

In this paper we use the semi-analytical approach to investigate the 
gravitational lensing of quasars by foreground dark halos. We work with
three kinds of cosmological models: SCDM, OCDM, and LCDM models. The mass 
function of dark halos is taken to be the
Press-Schechter function. The power spectrum of primordial fluctuations is
normalized by the cluster abundance constraints \citep{wan98,wan00}. We
alternatively consider three kinds of density profiles: SIS, NFW, and GNFW
profiles, where the last model allows us to estimate (by varying the free 
parameters in the GNFW profile) variants of the CDM scenario. For the 
various models we will calculate the lensing probability of splitting 
angles of multiple images, and test the sensitivity of the lensing 
probability to the input parameters regarding the properties of halos and
cosmological models.

The paper is organized as follows: In section \ref{section2} we outline the
cosmological models to be calculated in the paper, and introduce the power
spectrum of primordial fluctuations and the Press-Schechter function. In
section \ref{section3} the most simple case is investigated -- SIS halos
as lenses. We find that, the SCDM model produces fewer lensing events with 
large splitting angles, but more lensing events with small splitting 
angles than the OCDM model and the LCDM model. This is due to the fact
that for large splitting angles, which are produced by large mass halos, 
the Press-Schechter function is exponentially sensitive to the amplitude 
of primordial fluctuations and the SCDM model has the smallest amplitude
of primordial fluctuations; but for small splitting angles, which are 
produced by small mass halos, the Press-Schechter function is not 
sensitive to the amplitude of primordial fluctuations but the mean mass
density in the universe plays a more dominant role. In section 
\ref{section4} we consider the case for the NFW profile and the GNFW 
profile. As expected, we find that, the lensing probability is very 
sensitive to the mass density profile of the lenses. The more concentrated
toward center the density distribution is, the higher lensing probability
it produces. Compared to a SIS halo, a NFW halo is very ineffective in 
producing multiple images. For the NFW profile, the SCDM model produces 
the least lensing events: the OCDM model and the LCDM model produce a
comparable number of lensing events which exceed the number of lensing
events produced by the SCDM model by two orders of magnitudes. This is
mainly due to two facts: a NFW model in a SCDM universe has the smallest
concentration parameter, and a SCDM universe has the smallest amplitude of
fluctuations. In section \ref{section5}, we compare our predictions with 
observations. We find that, none of the models can explain the current 
observations: the SIS models predict too many lensing events with large 
splitting angles, while the NFW models predict too few lensing events 
with small splitting angles. Essentially, the observed high ratio of small
splitting to large splitting events is not predicted correctly. By a 
combination of them -- using the SIS profile for small mass halos and the 
NFW profile for large mass halos, we show that the observation results can 
be reasonably re-produced if we choose the mass at the transition point 
between the two halo populations to be $\sim 10^{13}$ solar masses -- the 
mass of halos below which cooling of the corresponding baryonic component 
will lead to concentration of the baryons to the inner parts of the mass 
profile \citep{ree77,blu86,por00}. Our results indicate that
the halos in the universe cannot be described by a single universal 
density profile, there exist at least two halo populations: small mass 
halos with a steep inner density slope (the inner parts dominated by the
baryonic component), and large mass halos with a shallow inner density 
slope (within which baryons and dark matter have essentially the same 
profiles). In section \ref{section6}, we draw our conclusions.

The failure of the SIS, and GNFW ($\alpha = 1.5$) models at large splitting 
angles appears to be
another manifestation of the tendency for CDM models to have too much power 
on small scales, i.e. too much mass concentration. From our sensitivity 
studies it appears that the cures proposed for other difficulties of 
CDM would help here as well, an example being the warm dark matter (WDM)
variant, for which $\alpha$ tends to be smaller \citep{bod00}. A model very 
roughly designed to match a WDM variant has a greatly reduced probability 
of large splitting angles: lower than the corresponding probability for the 
CDM model by one order of magnitude.

\section{Cosmological Models}
\label{section2}
The lensing probability is basically determined by three factors: (1) the 
positions of source objects; (2) the positions, the density profile, and 
the mass function of lenses; (3) the cosmological model. In this section 
we define the cosmological models that will be studied in this paper, and 
introduce the mass function of dark halos.

\subsection{Cosmological functions}
The cosmological models are classified with parameters $\Omega_m$, 
$\Omega_{\Lambda}$, and $\Omega_R$, which are respectively defined by 
\citep{pee93}
\begin{eqnarray}
    \Omega_m \equiv {\rho_0\over\rho_{\rm crit,0}}\,,
    \hspace{1 cm}
    \Omega_{\Lambda} \equiv {\Lambda\over 8\pi G\rho_{\rm crit,0}}\,,
    \hspace{1 cm}
    \Omega_R \equiv {-3k\over 8\pi G\rho_{\rm crit,0} a_0^2}\,,
\end{eqnarray}
where $\rho_0$ is today's average mass density (dark matter + baryonic 
matter + radiation) in the universe, $G$ is the gravitational constant,
$\Lambda$ is the cosmological constant, $a_0$ is today's scale of the 
universe, $k = 0, \pm 1$ is the spatial curvature of the universe, 
$\rho_{\rm crit,0}$ is today's critical mass density of the universe which 
is defined by
\begin{eqnarray}
     \rho_{\rm crit,0} \equiv {3H_0^2\over 8\pi G} \approx 1.88 \times 
        10^{-29}\, h^2\,{\rm g}\,{\rm cm}^{-3}\,,
\end{eqnarray}
where $H_0 = 100 h$ km s$^{-1}$ Mpc$^{-1}$ is the Hubble constant. The 
Einstein equations tell us that $\Omega_m + \Omega_{\Lambda} + \Omega_R = 
1$. Thus, among $\Omega_m$, $\Omega_{\Lambda}$, and $\Omega_R$, only two of 
them are independent parameters. In this paper we assume that the matter in 
the universe is dominated by cold dark matter (CDM), and we consider 
three kinds of universes:

1. {\it Einstein-de Sitter model} (SCDM). In this model, $\Omega_m = 1$,
$\Omega_{\Lambda} = \Omega_R = 0$, and $k = 0$. The linear growth function,
which describes the growth speed of the linear perturbation in the universe
\citep{pee80}, is
\begin{eqnarray}
     {\cal D}(z) = {1\over 1+z}\,,
\end{eqnarray}
where $z$ is the cosmic redshift. The linear growth function is normalized 
to be unity at present epoch: ${\cal D}(z=0)=1$.

2. {\it Open model} (OCDM). In an open universe, $0<\Omega_m<1$, 
$\Omega_{\Lambda} = 0$, $0<\Omega_R=1-\Omega_m<1$, and $k = -1$. The linear 
growth function is
\begin{eqnarray}
     {\cal D}(z) = {F_1\left({\Omega_m^{-1}-1\over 1+z}\right)\over
         F_1\left(\Omega_m^{-1}-1\right)}\,,
\end{eqnarray}
where
\begin{eqnarray}
     F_1(x)\equiv 1+{3\over x}+{3(1+x)^{1/2}\over x^{3/2}}\ln
        \left[(1+x)^{1/2}-x^{1/2}\right]\,.
\end{eqnarray}
$F_1$ is fitted by
\begin{eqnarray}
     F_1(x)\approx 1-\left({1.5\over x+1.5}\right)^{0.6}
     \label{a1}
\end{eqnarray}
with an error $< 1.3\%$ for $10^{-5}<x<10$.

3. {\it $\Lambda$-model} (LCDM). A $\Lambda$-universe has $0<\Omega_m<1$, 
$0<\Omega_{\Lambda}=1-\Omega_m<1$, $\Omega_R = 0$, and $k = 0$. The linear 
growth function is
\begin{eqnarray}
     {\cal D}(z) ={F_2\left[{(2 \Omega_{\Lambda}/\Omega_m)^{1/3}\over 
         1+z}\right]\over F_2\left[(2 \Omega_{\Lambda}/
	 \Omega_m)^{1/3}\right]}\,,
\end{eqnarray}
where
\begin{eqnarray}
     F_2(x)\equiv \left({x^3+2\over x^3}\right)^{1/2}\int_0^x
         \left({u\over u^3+2}\right)^{3/2} du\,.
\end{eqnarray}
$F_2$ is fitted by
\begin{eqnarray}
     F_2(x)\approx 0.358 \left[1-\left(1+0.23\, x^{2.2}\right)^{-1.26}
         \right]^{1/2.2}
     \label{a2}
\end{eqnarray}
with an error $< 2\%$ for $0<x<10^4$.

For all the three cosmological models, the proper cosmological distance 
from an object at redshift $z_1$ to an object at redshift $z_2$ along the 
same line of sight is \citep{pee93}
\begin{eqnarray}
     D\left(z_1,z_2\right) = {c\over H_0}\int_{z_1}^{z_2}{(1+z)^{-1} 
           dz\over \sqrt{\Omega_m(1+z)^3 + \Omega_R (1+z)^2 +
	   \Omega_{\Lambda} }}\,,
     \label{prop_dis}
\end{eqnarray}
where $c$ is the speed of light. In studying gravitational lensing, another
useful distance is the angular-diameter distance. The angular-diameter 
distance from an object at redshift $z_1$ to an object at redshift $z_2$ 
along the same line of sight is \citep{bar01}
\begin{eqnarray}
     D^A\left(z_1,z_2\right) = {c/H_0\over 1+ z_2}\,
            \left\{\begin{array}{ll}\left\vert\Omega_R\right\vert^{-1/2} 
	    \sin\left[\left\vert\Omega_R\right\vert^{1/2} w\right]
	    \hspace{0.5cm} & (k=1)\,,\\
            w & (k=0)\,,\\
            \Omega_R^{-1/2} \sinh\left[\Omega_R^{1/2} w\right] & 
	    (k=-1)\,,
   \end{array}
   \right.
   \label{ang_dis}
\end{eqnarray}
where
\begin{eqnarray}
     w\left(z_1,z_2\right) \equiv \int_{z_1}^{z_2}{dz \over \sqrt{
            \Omega_m(1+z)^3 + \Omega_R (1+z)^2 +\Omega_{\Lambda} }}\,.
\end{eqnarray}
Note that $D\left(z_1,z_2\right) = D\left(z_2,z_1\right)$ and $D\left(z_1,
z_2\right) = D\left(0,z_2\right) - D\left(0,z_1\right)$, but $D^A\left(z_1,
z_2\right) \neq D^A\left(z_2,z_1\right)$ and $D^A\left(z_1,z_2\right) \neq 
D^A\left(0,z_2\right) - D^A \left(0,z_1\right)$. But we have 
\begin{eqnarray}
     \left(1+z_2\right)D^A\left(z_1,z_2\right) = 
         \left(1+z_1\right)D^A\left(z_2,z_1\right)\,.
\end{eqnarray}

\subsection{Mass fluctuations}
Dark halos of galaxies and galaxy clusters are formed from primordial 
fluctuations of matter in the early universe. For a fluctuation with a 
power spectrum $P_k=|\delta_k|^2$, its variance on a scale of comoving 
radius $r$ is \citep{kol90}
\begin{eqnarray}
     \Delta^2(r)\equiv\left({\delta M\over M}\right)^2 =(2\pi)^{-3}
          \int_0^\infty 4\pi k^2 dk P_k W^2(kr)\,,
\end{eqnarray}
where $W(kr)$ is the Fourier transformation of a window function. For a 
top-hat window function
\begin{eqnarray}
     W(kr)=3\left[{\sin kr\over (kr)^3} - {\cos kr\over
          (k r)^2}\right]\,,
\end{eqnarray}
for a Gaussian window function
\begin{eqnarray}
     W(kr)= \exp[-k^2r^2/2]\,.
\end{eqnarray}
In this paper we use the top-hat window function. We compute the CDM power 
spectrum using the fitting formulae given by \citet{eis99}
\begin{eqnarray}
     {k^3\over 2\pi^2} P(k,z) = \delta_H^2 \left({ck\over 
         H_0}\right)^{3+n} T^2(k)\, {\cal D}(z)^2\,,
     \label{cdm_pow}
\end{eqnarray}
where $\delta_H$ is the amplitude of perturbations on the horizon scale 
today, $n$ is the initial power spectrum index, ${\cal D}(z)$ is the liner 
growth function, and
\begin{eqnarray}
     T &=& {L\over L + C q_{eff}^2}\,,
\end{eqnarray}
where
\begin{eqnarray}
     L &\equiv& \ln \left(e + 1.84 q_{eff}\right)\,, \hspace{1cm}
           q_{eff}\equiv {k\over \Omega_m h^2\, {\rm Mpc}^{-1}}\,,\\
     C &\equiv& 14.4 + {325\over 1+ 60.5 q_{eff}^{1.11}}\,.
\end{eqnarray} 
We normalize the power spectrum to $\sigma_8^2 \equiv \Delta^2\left(z=0,
r = 8 h^{-1} {\rm Mpc}\right)$, then
\begin{eqnarray}
     \delta_H = {\sigma_8 \over \left[\int_0^{\infty} {dk\over k} 
            \left({ck\over H_0}\right)^{3+n} T^2 \left(k\right) W^2 
            \left(k r_8\right)\right]^{1/2}}\,.
\end{eqnarray}
The value of $\sigma_8$ can be estimated from the cluster abundance 
constraints \citep{wan98,wan00}
\begin{eqnarray}
     \sigma_8 \Omega_m^\gamma \approx 0.5\,,
     \label{s8w}
\end{eqnarray}
where $\gamma \approx 0.43 + 0.33 \Omega_m$ for LCDM and $\gamma \approx 
0.33 + 0.35 \Omega_m$ for OCDM. For a scale-invariant spectrum predicted 
by the inflation theory and being consistent with observed cosmic 
microwave background (CMB) fluctuations, we have $n=1$ \citep{kol90}.

\subsection{Mass function of dark halos}
Assume the primordial density fluctuations are Gaussian. Then, according
to the Press-Schechter theory \citep{pre74}, the comoving number density 
of dark halos formed by redshift $z$ with mass in the range $(M,M+dM)$ is
\begin{eqnarray}
     n(M,z)\,dM = {\rho_0\over M}\, f(M,z)\, dM\,,
\end{eqnarray}
where $\rho_0\equiv\Omega_m\,\rho_{{\rm crit},0}$ is the present mean mass
density in the universe, $f(M,z)$ is the Press-Schechter function
\begin{eqnarray}
     f(M,z) = -\sqrt{2\over\pi}\,{\delta_c(z)\over M\Delta}\,
             {d\ln\Delta\over d\ln M}\,\exp\left[-{\delta_c^2(z)\over 
	     2\Delta^2}\right]\,,
     \label{p-s}
\end{eqnarray}
where $\Delta^2 = \Delta^2 (M,z=0)$ is the present variance of the 
fluctuations in a sphere containing a mean mass $M$, and $\delta_c (z)$ is 
the density threshold for spherical collapse by redshift $z$. The 
redshift-dependent density threshold has been calculated by many people 
\citep[and references therein]{lac93,koc95,eke96}, here we use the 
approximation \citep{nav97}
\begin{eqnarray}
     \delta_c(z) = {\delta_c^0[\Omega(z)] \over {\cal D}(z)}\,,
     \label{dec}
\end{eqnarray}
where
\begin{eqnarray}
     \Omega(z) \equiv {\rho \over \rho_{\rm crit}} = {\Omega_m (1+z)^3
         \over \Omega_m (1+z)^3 + \Omega_R (1+z)^2 + \Omega_\Lambda}\,,
\end{eqnarray}
${\cal D}(z)$ is the linear growth function normalized to unity at $z = 0$,
and
\begin{eqnarray}
    \delta_c^0 (\Omega) \approx 1.6865\times\left\{\begin{array}{ll}
       1, & \mbox{if $\Omega_m = 1$ and $\Lambda=0$}\\
       \Omega^{0.0185}, & \mbox{if $\Omega_m<1$ and $\Lambda=0$}\\
       \Omega^{0.0055}, & \mbox{if $\Omega_m + \Omega_{\Lambda} =1$}
       \end{array}
       \right.\,.
\end{eqnarray}
To a very good approximation $\delta_c(z) {\cal D}(z) \approx 1.69$ for 
all cosmological models with $\Omega_m > 0.1$.

In equation (\ref{p-s}), $r$ is related to $M$ by
\begin{eqnarray}
     M = {4\pi\over3}\rho_0 r^3 
       = {4\pi\over3}\Omega_m\,\rho_{{\rm crit},0}\, r^3\,, 
\end{eqnarray}
i.e.
\begin{eqnarray}
     r = 9.510 \left({1\over\Omega_m}{M\over 10^{15} h^{-1}
         M_{\odot}}\right)^{1/3} h^{-1} {\rm Mpc}\,.
\end{eqnarray}

Though it is obtained from very simple considerations, the Press-Schechter
function has been shown to be in remarkable agreement with N-body 
simulations for the standard cold dark matter scenario \citep{efs90,kau93} 
and observations \citep{bah93,gir98,gir00}.

The fluctuation variance $\Delta^2$ decreases with increasing mass $M$.
Thus, from equation (\ref{p-s}), $f(M,z)$ decreases exponentially with
increasing $M$. Because of this, a crude knowledge of $f(M,z)$ at the large
mass end gives a strict constraint on cosmological parameters 
\citep{chi98}. To see this, let us write the Press-Schechter function in 
the form
\begin{eqnarray}
    f(M,z,\sigma_8) = {f_1(M,z)\over\sigma_8}\,
       \exp\left[-{A(M,z)\over\sigma_8^2}\right]\,,
    \label{p-s1}
\end{eqnarray}
where we have used the fact that $\Delta\propto\sigma_8$. From equation 
(\ref{p-s1}), we have
\begin{eqnarray}
    {\delta f\over f} = {\delta\sigma_8\over \sigma_8}\,
        \left({2A\over\sigma_8^2}-1\right)\,.
\end{eqnarray}
So, if $A/\sigma_8^2\gg 1/2$ -- which is true for large $M$ -- a large 
error in $f$ corresponds to a small error in $\sigma_8$. As a simple 
example, consider a power law spectrum $P_k\propto k^{n_1}$, then we have
\begin{eqnarray}
    \Delta^2 (M) = \sigma_8^2\left[1.189\left({1\over\Omega_m}\,{M\over
    10^{15}h^{-1}M_{\sun}}\right)^{1/3}\right]^{-n_1-3}\,,
    \label{dem}
\end{eqnarray}
and
\begin{eqnarray}
    {2A\over\sigma_8^2} \propto M^{n_1+3\over 3} D(z)^{-2}\sigma_8^{-2}\,.
\end{eqnarray}
For large $M$ and large $z$, a large $\delta f/f$ corresponds to a small
$\delta\sigma_8/\sigma_8$ if $n_1 > -3$.

In the upper panel of Fig. \ref{fig1} we plot the Press-Schechter function
$f$ against the velocity dispersion $\sigma_v$ of dark halos at $z=0$ in
a SCDM cosmology, where the CDM power spectrum is given by equation
(\ref{cdm_pow}). The velocity dispersion of a dark halo is defined by
\begin{eqnarray}
    \sigma_v \equiv \left({G M\over 2 r_{200}}\right)^{1/2}\,,
    \label{sigv}
\end{eqnarray}
where $r_{200}$ is the radius of a sphere around a dark halo within which 
the average mass density is $200$ times the critical mean mass density of 
the universe, i.e. 
\begin{eqnarray}
    {M\over {4\pi\over 3}r_{200}^3} = 200 \rho_{\rm crit}\,,
    \label{r200}
\end{eqnarray}
and $M \equiv M_{200}$ is the mass within that sphere
\begin{eqnarray}
    M \equiv M_{200} = 4\pi\int_0^{r_{200}} \rho r^2 dr\,.
    \label{mass}
\end{eqnarray}
From equation (\ref{sigv}) and equation (\ref{r200}), we have
\begin{eqnarray}
    M &=& {\sigma_v^3\over G}\left({3\over 100\pi G
          \rho_{\rm crit}}\right)^{1/2} \nonumber\\
      &=& 0.656\times 10^{15} h^{-1} M_{\odot}\, \sigma_{v,1000}^3
        \left[\Omega_m(1+z)^3 + \Omega_R (1+z)^2 +
        \Omega_{\Lambda}\right]^{-1/2}\,,
      \label{mv}
\end{eqnarray}
where $M_{\odot}$ is the solar mass, $\sigma_{v,1000}\equiv{\sigma_v\over 
1000{\rm km/s}}$, and we have used
\begin{eqnarray}
    \rho_{\rm crit}(z) = \rho_{\rm crit,0}\,{\Omega_m\over\Omega(z)}\,
       \left(1+z\right)^3 = \rho_{\rm crit,0}\left[\Omega_m\left(1+
       z\right)^3+\Omega_R\left(1+z\right)^2+\Omega_{\Lambda}\right]\,.
    \label{rhoz}
\end{eqnarray}
In the upper panel of Fig. \ref{fig1}, the solid curve is for the case of
$\sigma_8=0.5$, the dashed curve is for the case of $\sigma_8=0.6$. It
clearly shows that with the same amount of change in $\sigma_8$, a larger 
change in $f$ happens at the end of large $\sigma_v$. In the lower panel 
of Fig. \ref{fig1}, we plot $\zeta\equiv{\delta f/f\over\delta\sigma_8/
\sigma_8}$ against $\sigma_v$ for a SCDM cosmology with $z=0$ and 
$\sigma_8=0.5$. Again, it shows that for large $\sigma_v$ a poor knowledge
in $f$ gives a good estimation of $\sigma_8$. This is true because the 
number density of halos with high velocity dispersion (which is true for 
rich clusters of galaxies) is very sensitive to the value of $\sigma_8$.
Thus, since lensing is produced primarily by the highest velocity 
dispersion clusters, the observed number of lenses will sensitively 
constrain the value of $\sigma_8$.

\section{Lensing by a Singular Isothermal Sphere}
\label{section3}
\subsection{Singular isothermal sphere as a lens}
The most simple model for a lens is a singular isothermal sphere (SIS) 
with a mass density
\begin{eqnarray}
    \rho(r) = {\sigma_v^2\over 2\pi G}\,{1\over r^2}\,,
    \label{rho}
\end{eqnarray}
where $\sigma_v$ is the velocity dispersion \citep{got74,tur84,sch92}. 
Though it is simple, this model can describe the flat rotation curves of 
galaxies and many basic features of gravitational lensing. Though the 
gravitational lensing by a SIS has been well studied in the literature
\citep{got74,tur84,nar88,koc95}, we present it here since we are using an 
updated CDM power spectrum and we want to compare its results with those
for the NFW and the GNFW cases.

The surface mass density of the SIS is
\begin{eqnarray}
    \Sigma(\xi) = {\sigma_v^2\over 2G}\,{1\over\xi}\,,
    \label{sig}
\end{eqnarray}
where $\xi\equiv|\vec{\xi}|$, $\vec{\xi}$ is the position vector in the 
lens plane. Choose the length scales in the lens plane and the source 
plane to be respectively
\begin{eqnarray}
    \xi_0 = 4\pi\left({\sigma_v\over c}\right)^2\,
          {D^A_L D^A_{LS}\over D^A_S}\,,
    \hspace{1 cm}
    \eta_0 = \xi_0\,{D^A_S\over D^A_L}\,,
    \label{xi0}
\end{eqnarray}
where $D^A_S$ is the angular-diameter distance from the observer to the 
source object, $D^A_L$ is the angular-diameter distance from the observer 
to the lens object, $D^A_{LS}$ is the angular-diameter distance from the 
lens to the source object. Remember that $D^A_{LS} \neq D^A_S - D^A_L$.
Then the position vector of a point in the lens plane can be written as 
$\vec{\xi} = \vec{x} \xi_0$, the position vector of a point in the source 
plane can be written as $\vec{\eta} = \vec{y} \eta_0$. The lensing 
equation is
\begin{eqnarray}
    y = x - {m(x)\over x}\,,
    \label{lens1}
\end{eqnarray}
where
\begin{eqnarray}
    m(x)\equiv 2\int_0^x{\Sigma\left(x^\prime\right)\over\Sigma_{\rm 
           cr}}x^\prime dx^\prime\,,
    \label{mx}
\end{eqnarray}
where the critical surface mass density $\Sigma_{\rm cr}$, defined by
\citep{tur84}
\begin{eqnarray}
    \Sigma_{\rm cr} \equiv {c^2\over 4\pi G}\,{D^A_S\over D^A_L 
         D^A_{LS}}\,, \label{scr}
\end{eqnarray}
is of the order of the surface density of the universe.

Inserting equation (\ref{sig}) and equation (\ref{scr}) into equation
(\ref{mx}), then into equation (\ref{lens1}), we obtain ${\Sigma/
\Sigma_{\rm cr}} = (2|x|)^{-1}$ and
\begin{eqnarray}
    y = x - {|x|\over x}\,.
\end{eqnarray}
Double images are formed if and only if $|y|\le 1$, i.e. $|x|\le 1$. The 
separation between the two images is
\begin{eqnarray}
    \Delta x = 2,
    \label{dex}
\end{eqnarray}
thus the splitting angle is
\begin{eqnarray}
    \Delta\theta = {\xi_0\over D^A_L} \Delta x = 8\pi\left({\sigma_v
        \over c}\right)^2 {D^A_{LS}\over D^A_S} = 0.9613^\prime\,
        \sigma_{v,1000}^2\, {D^A_{LS}\over D^A_S}\,.
    \label{ds1}
\end{eqnarray}
Using equation (\ref{mv}), we have
\begin{eqnarray}
    \Delta\theta = {8\pi\over c^2}\,\left({100\pi\over 3}\,G^3 M^2
        \rho_{\rm crit}\right)^{1/3}\,{D^A_{LS}\over D^A_S} = 
	1.271^{\prime}\,\,{D^A_{LS}\over D^A_S}\,\left({M\over 
	10^{15}h^{-1} M_{\sun}}\right)^{2/3} \left({\rho_{\rm crit}
	\over\rho_{\rm crit,0}}\right)^{1/3}\,.
    \label{ds2}
\end{eqnarray}

The cross-section (defined in the lens plane) for forming two images with
splitting angle $\Delta\theta > \Delta\theta_0$ is
\begin{eqnarray}
    \sigma = \pi \xi_0^2\, \vartheta\left(\Delta\theta -
             \Delta\theta_0\right)
           = \pi \xi_0^2\, \vartheta\left(M-M_0\right)\,,
    \label{sig1}
\end{eqnarray}
where $\vartheta$ is the step function
\begin{eqnarray}
    \vartheta\left(u-v\right) = \left\{\begin{array}{ll}
       1, & \mbox{if $u>v$}\\
       0, & \mbox{if $u<v$}
       \end{array}
       \right.\,,
\end{eqnarray}
and
\begin{eqnarray}
    M_0 &=& \left({3\over 100\pi}\right)^{1/2}\,{c^3\over G^{3/2}
         \rho_{\rm crit}^{1/2}}\,\left({\Delta\theta_0\over
	 8\pi}\right)^{3/2}\,\left({D^A_S\over D^A_{LS}}\right)^{3/2}
	 \nonumber \\
        &=& 0.6975 \times 10^{15} h^{-1} M_{\sun}\,\left({\Delta
	\theta_0\over 1^{\prime}}\right)^{3/2}\,\left({D^A_S\over
	D^A_{LS}}\right)^{3/2}\,\left({\rho_{\rm crit}\over \rho_{\rm
	crit,0}}\right)^{-1/2}\,.
    \label{m0}
\end{eqnarray}

\subsection{Lensing probability by a singular isothermal sphere}
The integral lensing probability for a source object at redshift $z_s$
is \citep{sch92}
\begin{eqnarray}
    P = \int_0^{z_s}{dP\over dz} dz\,,
    \label{ip}
\end{eqnarray}
where $z$ is the redshift of lenses, and
\begin{eqnarray}
    {dP\over dz} &=& {d D_L \over dz} \int_0^{\infty}
                     \overline{n}(M,z)\sigma(M,z) dM  \nonumber \\
                 &=& \rho_{{\rm crit},0}\Omega_m (1+z)^3\,
		     {d D_L\over dz}\int_0^{\infty}{1\over M} f(M,z)
                     \sigma(M,z)\ dM\,,
    \label{dpz}
\end{eqnarray}
where $ \overline{n}(M,z) dM \equiv n(M,z)(1+z)^3 dM$ is the physical 
number density of dark halos of masses between $M$ and $M+dM$, $n(M,z) dM$
is the comoving number density of dark halos of masses between $M$ and $M+
dM$, $\sigma(M,z)$ is the lensing cross-section for a dark halo of mass 
$M$ at redshift $z$, $f(M,z)$ is the mass function of dark halos. Here 
$D_L$ is the proper distance from the observer to the lens object.

If we use $10^{15} h^{-1} M_{\odot}$ as the unit of $M$, then the unit of
$f(M,z)$ is $\left(10^{15} h^{-1} M_{\sun}\right)^{-1}$. Define $M_{15} 
\equiv M /\left(10^{15} h^{-1} M_{\sun}\right)$. Let us use $c/H_0 = 
2997.9\, h^{-1} {\rm Mpc}$ as the unit for cosmological distances, and 
$ h^{-1} {\rm Mpc}$ as the unit for local lengths (thus the unit of 
$\sigma$ is $ h^{-2} {\rm Mpc}^2$), then
\begin{eqnarray}
     {dP\over dz} = 0.8321\,\Omega_m (1+z)^3 {d\over dz}\left({D_L\over
          c/H_0}\right)\int_0^{\infty} f(M_{15},z) \left[{\sigma(M_{15},
	  z)\over 1 h^{-2}{\rm Mpc}^2}\right]{dM_{15}\over M_{15}}\,.
     \label{dpz1}
\end{eqnarray}

Inserting equation (\ref{sig1}) and equation (\ref{xi0}) into equation 
(\ref{dpz}), we have for the SIS case
\begin{eqnarray}
    {d\over dz} P(>\Delta\theta_0) = 16\pi^3 \rho_{{\rm crit},0}\Omega_m 
	 (1+z)^3\, {d D_L\over dz}\left({D^A_L D^A_{LS}\over 
	 D^A_S}\right)^2 \int_{M_0}^{\infty}{1\over M} f(M,z)
	\left({\sigma_v\over c}\right)^4 dM\,,
    \label{dpz2}
\end{eqnarray}
where $\sigma_v/c$ is related to $M$ by the reverse of equation 
(\ref{mv}), i.e.
\begin{eqnarray}
    {\sigma_v\over c} &=& \left({M\over M_1}\right)^{1/3}\left[\Omega_m
         (1+z)^3 +\Omega_R(1+z)^2+\Omega_{\Lambda}\right]^{1/6} 
	    \nonumber\\
	 &=& 3.836\times 10^{-3} M_{15}^{1/3}\left[\Omega_m(1+z)^3+
	     \Omega_R (1+z)^2+\Omega_{\Lambda}\right]^{1/6}\,,
    \label{sigv1}
\end{eqnarray}
where
\begin{eqnarray}
    M_1 \equiv \left({3c^6\over 100\pi G^3\rho_{\rm crit,0}}\right)^{1/2}
        = 1.772\times 10^{22}\,h^{-1}M_{\sun}\,.
\end{eqnarray}
Or, from equations (\ref{sig1}), (\ref{xi0}), and (\ref{dpz1}), we have
\begin{eqnarray}
    {d\over dz} P(>\Delta\theta_0) &=& 0.4593\, \Omega_m (1+
	 z)^3\left({D^A_R\over c/H_0} \right)^2\,{d\over dz}\left({D_L
	 \over c/H_0}\right)\nonumber\\ 
         &&\times\int_{M_0}^{\infty} f(M_{15},z) 
	 \left({\sigma_v\over 10^3{\rm km/s}}\right)^4{dM_{15}
	 \over M_{15}}\,,
    \label{dpz3}
\end{eqnarray}
where $D^A_R \equiv D^A_{L} D^A_{LS} / D^A_S$\,.

With equations (\ref{ip}), (\ref{sigv1}), and (\ref{dpz3}), we can 
calculate $dP(>\Delta\theta_0)/dz$ and $P(>\Delta\theta_0)$ for a given 
cosmology (specified by $\Omega_m$, $\Omega_{\Lambda}$, and the Hubble 
constant $h$), a given primary perturbation (specified by $\sigma_8$), and 
a given position of the source object ($z_s$), assuming the CDM power 
spectrum is given by equation (\ref{cdm_pow}) and the mass function of 
dark halos is given by equation (\ref{p-s}). In our numerical calculations 
we assume $h=0.7$, $z_s=1.5$, and $\sigma_8$ is related to $\Omega_m$ by 
equation (\ref{s8w}). In Fig. \ref{fig2} and Fig. \ref{fig3} we show the
results for three typical cosmological models: SCDM with $\Omega_m = 1$ 
and $\sigma_8=0.5$; OCDM with $\Omega_m = 0.3$, $\Omega_\Lambda = 0$, and 
$\sigma_8 = 0.85$; and LCDM with $\Omega_m = 0.3$, $\Omega_\Lambda = 0.7$, 
and $\sigma_8 = 0.95$.

Fig. \ref{fig2} shows the differential lensing probability $dP/dz$ for
$\Delta\theta > \Delta\theta_0 = 5^{\prime\prime}$. The maximum of 
$dP/dz$ is at $z\approx 0.3$ for SCDM, at $z\approx 0.4$ for OCDM and 
LCDM. Median expected redshifts for the three models are respectively 
$0.349$, $0.455$, and $0.461$. This is consistent with the demonstration 
that the differential 
lensing probability peaks at intermediate redshifts (i.e. roughly the half 
way to the source object; Turner, Ostriker, \& Gott 1984) and with the 
scenario that structures form later in the SCDM model than in OCDM and 
LCDM models \citep{lon98,bar98}. As is well known structure forms earlier 
in low $\Omega_m$ universes, so the relative probability of a lens being 
at $z=0.1$ rather than $1.0$ is orders of magnitude higher in SCDM than
it is in OCDM/LCDM. Fig. 3 shows the
integral probability $P(>\Delta\theta)$ as a function of the splitting
angle $\Delta\theta$. For small $\Delta\theta$, the SCDM model produces
more lensing events than the OCDM and the LCDM models, which is a
manifestation of the fact that the SCDM universe has a higher mean mass
density than the OCDM and the LCDM universes. For large $\Delta\theta$,
the SCDM model produces less lensing events than the OCDM and the LCDM
models, which is caused by the fact that at large mass end the
Press-Schechter function is very sensitive to $\sigma_8$ (see Fig.
\ref{fig1}) while the SCDM model has a much smaller $\sigma_8$ than the
OCDM and the LCDM models. The difference in the lensing probabilities for
the OCDM model and the LCDM model is mainly due to the effect of the
cosmological constant: the cosmological constant tends to increase the
lensing probability \citep{tur90}, primarily because there is a greater
metric distance and thus more potential lenses between an observer and
a source with a given redshift in $\Lambda$ cosmologies.

\section{Lensing by a Generalized NFW Model}
\label{section4}
\subsection{The generalized NFW model as a lens}
\label{section4.1}
The SIS model is simple and can describe many observational features of 
galaxies and galaxy clusters. However, the SIS does not fit well the 
density profiles predicted by N-body numerical simulations 
\citep{fre85,fre88,qui86,efs88,zur88,war92,cro94}. \citet{nav96,nav97}
proposed a ``universal'' mass density profile for dark halos: the 
so-called NFW profile, which fits the simulation results better than the 
SIS. The NFW profile is
\begin{eqnarray}
    \rho_{\rm NFW}(r) = {\rho_s r_s^3 \over r(r+r_s)^2}\,,
    \label{nfw}
\end{eqnarray}
where $\rho_s$ and $r_s$ are constants. At large radii ($r \gg r_s$) 
$\rho_{\rm NFW}\sim r^{-3}$, at small radii ($r\ll r_s$) $\rho_{\rm NFW}
\sim r^{-1}$, both are different from the SIS. At intermediate radii the 
NFW profile resembles the SIS. 
Though Navarro, Frenk, \& White demonstrated that over two decades in 
radius their profile accurately fits halos with mass spanning about four 
orders of magnitude ranging from dwarf galaxy halos to those of rich 
galaxy clusters, with higher resolution simulations \citet{jin00} have 
argued that the NFW profile is not correct on small scales. Furthermore, 
\citet{sub00} and \citet{ghi00} have emphasized the dependence of the
inner slope of halo mass density on the form of the power spectrum of the 
primordial fluctuations. Therefore, here we consider a generalized NFW 
(GNFW) profile \citep{zha96}
\begin{eqnarray}
    \rho(r) = {\rho_s r_s^3 \over r^{\alpha}(r+r_s)^{3-\alpha}}\,,
    \label{gnfw}
\end{eqnarray}
where $\alpha$ ($0<\alpha<3$) is a new constant parameter. The NFW profile 
is a specific case of the GNFW profile when $\alpha = 1$. Obviously, at
large radii ($r \gg r_s$) the GNFW profile has the same behavior as the 
NFW profile. But on small scales ($r\ll r_s$) they are different unless 
$\alpha = 1$. When $\alpha = 2$, the GNFW profile resembles an SIS on 
small scales but resembles the NFW profile on large scales.

The surface mass density for the GNFW profile is
\begin{eqnarray}
    \Sigma(x) = 2\rho_s r_s\int_0^{\infty}\left(x^2+z^2\right)^{-\alpha/2}
       \left[\left(x^2+z^2\right)^{1/2}+1\right]^{-3+\alpha} dz\,,
    \label{sur_nfw}
\end{eqnarray}
where $x=|\vec{x}|$ and $\vec{x} = \vec{\xi}/r_s$, $\vec{\xi}$ is the 
position vector in the lens plane. Inserting equation (\ref{sur_nfw}) 
into equation (\ref{lens1}) and equation (\ref{mx}), we obtain the lensing 
equation for a GNFW halo
\begin{eqnarray}
    y = x - \mu_s {g(x)\over x}\,,
    \label{le1}
\end{eqnarray}
where $y = |\vec{y}|$, $\vec{\eta} = \vec{y}\, r_s D^A_S/D^A_L$ is the 
position vector in the source plane, and
\begin{eqnarray}
    g(x) \equiv \int_0^x u du \int_0^{\infty} \left(u^2 + 
           z^2\right)^{-\alpha/2} \left[\left(u^2+z^2\right)^{1/2} + 
	   1\right]^{-3+\alpha} dz\,,
    \label{gx}
\end{eqnarray}
and
\begin{eqnarray}
    \mu_s \equiv {4\rho_s r_s\over \Sigma_{\rm cr}}\,,
    \label{alps}
\end{eqnarray}
where $\Sigma_{\rm cr}$ is the critical surface mass density defined by 
equation (\ref{scr}). Note, here we use $\xi_0=r_s$ as the length unit in 
the lens plane, $\eta_0 = r_s D^A_S/D^A_L$ as the length unit in the 
source plane. The dimensionless parameter $\mu_s$ summarizes the ability 
for a GNFW halo to produce multiple images. Multiple images are formed 
if and only if $|y|\leq y_{\rm cr}$, where $y_{\rm cr} \equiv - 
y(x_{\rm cr})$, $x_{\rm cr}>0$ is determined by $dy/dx = 0$ (see Fig. 
\ref{fig4}). Let us consider the splitting angle $\Delta \theta$ between 
the two outside images when more than two images are formed. For $|y| < 
y_{\rm cr}$, there are three real roots of equation (\ref{le1}): $x_1>x_2
>x_3$, then $\Delta\theta\propto \Delta x \equiv x_1-x_3$. In general, the 
value of $x_1-x_3$ (and thus the value of $\Delta\theta$) is insensitive 
to the value of $y$ when $|y|< y_{\rm cr}$~\footnote{See \citet{sch92}.
For the extreme case of a singular isothermal sphere, $\Delta x$ is 
exactly independent the position of the source object in the source plane 
when double images are formed, see equation [\ref{dex}].}. So, we have
\begin{eqnarray}
    \Delta x\,(y) \approx \Delta x\,(y=0) = 2 x_0\,, \hspace{1 cm}
         {\rm for}\,\,|y|<y_{\rm cr}\,,
\end{eqnarray}
where $x_0$ is the positive root of $y(x) = 0$. Then, for a GNFW lens at 
redshift $z$, the cross-section in the lens plane for forming multiple 
images with $\Delta\theta>\Delta\theta_0$ is
\begin{eqnarray}
    \sigma\left(>\Delta\theta_0,M,z\right) \approx \pi y_{\rm cr}^2 
        r_s^2\,\vartheta\left(\Delta\theta - \Delta\theta_0\right)\,.
    \label{signfw}
\end{eqnarray}
The splitting angle $\Delta\theta$ is given by
\begin{eqnarray}
    \Delta\theta = {r_s\over D^A_L} \Delta x \approx
        {2 x_0 r_s\over D^A_L}\,.
    \label{deth}
\end{eqnarray}

For $\alpha = 1$ (the NFW case) and $\alpha=2$ (the modified SIS case), 
$g(x)$ defined by equation (\ref{gx}) can be worked out analytically. For 
$\alpha = 1$, we have
\begin{eqnarray}
    g(x) = \ln{x\over 2} +
          \left\{\begin{array}{ll}
          {1\over \sqrt{x^2-1}} \arctan\sqrt{x^2-1}
          \hspace{1 cm} &(x>1)\,,\\
          {1\over \sqrt{1-x^2}}\, {\rm arctanh}\sqrt{1-x^2}
          \hspace{1 cm} &(0<x<1)\,,\\
          1 \hspace{1cm &}(x=1)\,.
   \end{array}
   \right.
   \label{gx1}
\end{eqnarray}
For $\alpha = 2$, we have
\begin{eqnarray}
    g(x) = \ln{x\over 2} + {\pi\over 2}x +
          \left\{\begin{array}{ll}
          -\sqrt{x^2-1}\, \arctan\sqrt{x^2-1}
          \hspace{1 cm} &(x>1)\,,\\
          \sqrt{1-x^2}\, {\rm arctanh}\sqrt{1-x^2}
          \hspace{1 cm} &(0<x<1)\,,\\
          1 \hspace{1cm &}(x=1)\,.
   \end{array}
   \right.
   \label{gx2}
\end{eqnarray}
Equation (\ref{gx1}) has also been obtained by \cite{bar96}. For other 
values of $\alpha$, $g(x)$ has to be worked out numerically.

\subsection{Determination of $\rho_s$ and $r_s$}
For a halo with a GNFW profile, its mass diverges logarithmically as
$r\rightarrow\infty$. So as usual, we define the mass of a halo to be the 
mass within $r_{200}$
\begin{eqnarray}
    M = 4\pi\int_0^{r_{200}}\rho r^2 dr = 4\pi \rho_s r_s^3
        f(c_1)\,,
    \label{gm200}
\end{eqnarray}
where $c_1 \equiv r_{200}/r_s$ is the concentration parameter and
\begin{eqnarray}
    f(c_1) \equiv \int_0^{c_1} {x^2 dx\over x^\alpha (1+x)^{3-\alpha}}\,.
\end{eqnarray}
For $1\le \alpha \le 2$, $f(c_1)$ can be worked out analytically
\begin{eqnarray}
    f(c_1) \equiv \left\{\begin{array}{ll}
          \ln(1+c_1)-{c_1\over 1+c_1} \hspace{1 cm} &(\alpha = 1)\,,\\
          \ln(1+c_1) \hspace{1 cm} &(\alpha = 2)\,,\\
          {c_1^{3-\alpha}\over3-\alpha}\,_2F_1(3-\alpha,3-\alpha;4-
	  \alpha;-c_1) \hspace{1cm &}(1<\alpha<2)\,,
   \end{array}
   \right.
   \label{fc}
\end{eqnarray}
where $_2F_1$ is the hypergeometric function. From equation (\ref{r200}) 
and equation (\ref{gm200}), we obtain
\begin{eqnarray}
    \rho_s = \rho_{\rm crit}\,{200\over 3}\,{c_1^3\over f(c_1)}=
             \rho_{\rm crit,0}\left[\Omega_m\,(1+z)^3+\Omega_R
             (1+z)^2+\Omega_{\Lambda}\right]{200\over 3}\,{c_1^3\over 
	     f(c_1)}\,,
    \label{rhos}
\end{eqnarray}
and
\begin{eqnarray}
    r_s = {1\over c_1}\left({3M\over 800\pi\rho_{\rm crit}}\right)^{1/3}
        = {1.626\over c_1}\,{M_{15}^{1/3}\over\left[\Omega_m\,(1+z)^3+
	   \Omega_R (1+z)^2+\Omega_{\Lambda}\right]^{1/3}}\,h^{-1} 
	   {\rm Mpc}\,.
    \label{rs}
\end{eqnarray}
If we know the value of the concentration parameter $c_1$, with equation
(\ref{rhos}) and (\ref{rs}) we can determine $\rho_s$ and $r_s$ for any 
halo of mass $M$ in any cosmology. Interestingly, $\rho_s$ does not depend 
on the mass of the halo. To determine the value of $c_1$ is not easy, and 
different methods give different results \citep{nav97,bar98}. Various
simulations suggest that $c_1(z) \propto (1+z)^{-1}$ \citep{nav00,bul01}.
Thus, here we assume that $c_1(z) = c_1(z=0)/ (1+z)$, and we try to infer 
$c_1(z=0)$ from simulation results. We take the values of $c_1(z=0)$ for 
the NFW profile (i.e. the $\alpha = 1$ case) from \citet{bar98}'s 
simulation results, then obtain the values of $c_1(z=0)$ for other cases 
(i.e. $\alpha> 1$) by referencing the values for the NFW case. To do so, 
let us
assume that fitting a dark halo with different density profiles gives the 
same ratio $\eta \equiv r_{1/2}/r_{200}$, where $r_{1/2}$ is defined by 
$M(r<r_{1/2}) = {1\over 2} M(r<r_{200})$. Then we obtain a relation
\begin{eqnarray}
    \int_0^{\eta c_1(z=0)}{x^2 dx\over x^\alpha(1+x)^{3-\alpha}} =
         {1\over 2}\int_0^{c_1(z=0)}{x^2 dx\over x^\alpha(1+x)^{3-
	 \alpha}}\,.
    \label{rel}
\end{eqnarray}
In equation (\ref{rel}), $\eta$ is the same for all values of $\alpha$. 
So, with the known value of $c_1$ for the $\alpha=1$ case -- let us denote 
it with $c_0$, we can solve $\eta = \eta(c_0)$ by equation (\ref{rel}). 
Then, since $\eta$ is assumed to be the same for all the values of 
$\alpha$, we can solve $ c_1(z=0) = c_1 (\alpha,z=0)$ by equation 
(\ref{rel}) for any $1<\alpha\le 2$ with the $\eta$ just solved. Then, we 
calculate $c_1$ at any redshift with $c_1(z) = c_1(z=0)/(1+z)$.

According to \citet{bar98}, for the NFW profile, we choose
\begin{eqnarray}
    c_1(z=0) = \left\{\begin{array}{ll}
       5, & \mbox{for SCDM}\\
       9, & \mbox{for OCDM}\\
       7, & \mbox{for LCDM}
       \end{array}
       \right.\,.
       \label{con_nfw}
\end{eqnarray}
Then, according to the procedure described above, we obtain
\begin{eqnarray}
    c_1(z=0) = \left\{\begin{array}{ll}
       2.7, & \mbox{for SCDM}\\
       5.3, & \mbox{for OCDM}\\
       4.0, & \mbox{for LCDM}
       \end{array}
       \right.
       \label{con_nfw2}
\end{eqnarray}
for a GNFW profile with $\alpha = 1.5$, and
\begin{eqnarray}
    c_1(z=0) = \left\{\begin{array}{ll}
       0.58, & \mbox{for SCDM}\\
       1.8, & \mbox{for OCDM}\\
       1.2, & \mbox{for LCDM}
       \end{array}
       \right.
       \label{con_nfw3}
\end{eqnarray}
for a GNFW profile with $\alpha = 2$. 

A critical parameter determining gravitational lensing is the surface mass 
density. A mass concentration with a central surface density larger than 
the surface density of the universe, $\Sigma_{\rm cr} \propto c H_0/G$ 
(see eq. [\ref{scr}]), can produce multiple images \citep{tur84}. While it 
is true that for $\alpha\ge 1$ the surface density is divergent as $r
\rightarrow 0$ and thus, formally, all halos can produce multiple images, 
there may be very little mass contained within the $\Sigma = \Sigma_{\rm 
cr}$ contour. So, large splittings will not be common unless the surface 
density at the half-mass point is near $\Sigma_{\rm cr}$. The mean surface
density at the half-mass radius is
\begin{eqnarray}
    \Sigma_{1/2} &=& {M\over 2\pi} \left({800 \pi \rho_{\rm crit}\over
        3 M}\right)^{2/3} {1\over \eta^2}\nonumber\\
          &=& 0.0126\, {{\rm g}\over{\rm cm}^2}\,
          M_{15}^{1/3} h \left[\Omega_m (1+z)^3 + \Omega_R (1+z)^2
          + \Omega_{\Lambda}\right]\, S(c_1)\,,
    \label{shalf}
\end{eqnarray}
where $S(c_1)\equiv\eta(c_1)^{-2}$. For a fixed cosmology and a fixed mass
$\Sigma_{1/2}$ is proportional to $S(c_1)$. We plot this function in Fig. 
\ref{fig5}. The steep dependence of $S(c_1)$ ($\propto\Sigma/\Sigma_{\rm 
cr}$) on $c_1$ indicates that, for any assumed $\alpha$, more concentrated 
halos are much more effective in lensing.

Several authors have noted that the large concentrations indicated by high
resolution N-body simulations of the CDM scenario \citep{nav96,nav97,moo98} 
may be inconsistent
with a variety of observations including the inner rotation curves of 
galaxies \citep{moo94,flo94,blo97,tys98,spe00}. Here we note that if we 
alter, for whatever reasons, the concentration parameter from ten to five 
we will lower the characteristic half-mass surface density by a factor of 
$\sim 1.5$ reducing greatly the fraction of mass in the universe contained 
in halos with a mean surface density exceeding the critical surface 
density, and thus reducing the probability of lensing by a large factor.
This will be confirmed by the results in the next subsection. Recently, 
\citet{bod00} have performed detailed high-resolution 
N-body simulations of the warm dark matter (WDM) scenario to determine if 
this variant can successfully address the putative difficulties of the CDM 
paradigm. They find a significant decrease in concentration in the WDM
scenario and also a decrease in the best fit value of $\alpha$.

\subsection{Lensing probability by a generalized NFW model}
Once $\rho_s$ and $r_s$ are determined, we can calculate $\mu_s$ with
\begin{eqnarray}
    \mu_s = 2.0014\times 10^{-3} \left({\rho_s\over \rho_{{\rm crit},0}}
            \right)
            \left({r_s\over 1 h^{-1} {\rm Mpc}}\right)
            \left({D^A_R\over c/H_0}\right)\,,
    \label{alps1}
\end{eqnarray}
where equations (\ref{scr}) and (\ref{alps}) have been used. With equations
(\ref{le1}), (\ref{deth}), (\ref{rhos}), (\ref{rs}), and (\ref{alps1}), we
can solve $\Delta\theta = \Delta\theta (M,z)$ for any $1\le \alpha \le 2$.
In Fig. \ref{fig6} we show $\Delta\theta = \Delta\theta(M)$ produced by a
NFW lens at $z=0.3$, where the concentration parameters are given by 
equation (\ref{con_nfw}). The source object is assumed to be at $z_s= 1.5$.
For comparison, we also show the splitting angle produced by a SIS lens at
the same redshift (thin lines in Fig. \ref{fig6}). The difference in the 
splitting angles produced by the two different mass density profiles is 
dramatic for small mass lenses: the splitting produced by a NFW lens is 
greatly shifted towards small angles. To produce the same small splitting 
angle a NFW halo requires more mass than a SIS halo. And, for the NFW case 
the difference in the results for different cosmological models are 
important -- especially for low mass lenses, but for the SIS case the 
difference in the results for different cosmological models are 
unimportant.
 
The splitting angle $\Delta\theta$ increases monotonically with increasing 
$M$, so $\vartheta (\Delta\theta - \Delta\theta_0) = \vartheta (M - M_0)$, 
where $M_0$ is obtained by solving $\Delta\theta (M_0,z) = \Delta\theta_0$. 
Then, from equation (\ref{dpz1}) and equation (\ref{signfw}), the 
differential lensing probability by a GNFW profile is
\begin{eqnarray}
     {d\over dz} P(>\Delta\theta_0) &=& 2.614\,\Omega_m (1+z)^3\,
          {d\over dz}\left({D_L\over c/H_0}\right)\nonumber\\
          &&\times\int_{M_0}^{\infty} f\left(M_{15},z\right) 
          y_{\rm cr}^2 \left({r_s\over 1 h^{-1} {\rm Mpc}}\right)^2
          {dM_{15}\over M_{15}}\,.
     \label{dpznfw}
\end{eqnarray}
The integral lensing probability is calculated with equation (\ref{ip}) 
and equation (\ref{dpznfw}).

We have calculated the lensing probabilities for different cosmologies and 
different GNFW halos. Our results are summarized in Fig. \ref{fig7} -- 
Fig. \ref{fig10}, where the source object is assumed to be at $z_s = 1.5$,
the Hubble constant is taken to be $h = 0.7$, $\sigma_8$ and $\Omega_m$ are
constrained by equation (\ref{s8w}). 

Fig. \ref{fig7} shows the differential
lensing probability $dP/dz$ for $\Delta\theta>5^{\prime\prime}$ as a 
function of lenses' redshift $z$. The three cosmological models are: SCDM
with $\Omega_m = 1$ and $\sigma_8 = 0.5$; OCDM with $\Omega_m = 0.3$, 
$\Omega_\Lambda = 0$, and $\sigma_8 = 0.85$; LCDM with $\Omega_m = 0.3$,
$\Omega_\Lambda = 0.7$, and $\sigma_8 = 0.95$. Three cases with different
values of $\alpha$ are shown: $\alpha = 1$ (i.e. the NFW case), $1.5$, and 
$2$. The concentration parameters of halos are determined with the 
procedure described in the last subsection, i.e. they are given by equation
(\ref{con_nfw}), equation (\ref{con_nfw2}), and equation (\ref{con_nfw3}).
Similar to the case for SIS halos (see Fig. \ref{fig2}) and
independent of the parameter $\alpha$ of GNFW halos, the differential 
lensing probability peaks at intermediate redshifts: at $z \approx 0.3$ for 
SCDM, at $z \approx 0.4$ for OCDM and LCDM. Fig. \ref{fig8} shows the 
integral lensing probability $P(>\Delta\theta)$ as a function of the
splitting angle $\Delta\theta$, for the same models in Fig. \ref{fig7}.
As $\alpha$ decreases (i.e. as the halo central density becomes more 
shallow),
Fig. \ref{fig8} shows, the lensing probability drops quickly. NFW halos
are least efficient in producing multiple images among halos with
$1\le\alpha\le2$. Furthermore, as $\alpha$ decreases, the difference
between the SCDM cosmological model and the OCDM/LCDM model becomes more
prominent: for the NFW case the lensing probability for the SCDM model is
lower than the lensing probability for the OCDM/LCDM model by two orders
of magnitude, though the difference between the OCDM model and the LCDM
model is not so big. This is due to the fact that the lensing probability
for the NFW case is extremely sensitive to the concentration parameter of
halos and the mass fluctuation $\sigma_8$, a halo in a SCDM universe has
the smallest concentration parameter $c_1$ (see eq. [\ref{con_nfw}]), and
the SCDM model has the smallest $\sigma_8$. Fig. \ref{fig8} shows one of 
our most important results. If $\alpha$ is the same for all halos,
then one expects that, in the OCDM/LCDM case, the number of $10^{\prime
\prime}$ splittings observed to be only slightly less than the number of 
$1^{\prime\prime}$ splittings observed. 

The sensitivity of the lensing
probability to the concentration parameter is shown in Fig. \ref{fig9},
where the models are the same in Fig. \ref{fig7} and Fig. \ref{fig8}
except that in Fig. \ref{fig9} we allow the concentration parameter to
vary from $2$ to $15$. Fig. \ref{fig9} shows that the sensitivity of the
lensing probability to the concentration parameter increases quickly as
$\alpha$ and $c_1$ decrease. Fig. \ref{fig10} shows the dependence of the 
lensing probability $P(>5^{\prime\prime})$ on the cosmic mass density 
$\Omega_m$ when $\Omega_m$ and $\sigma_8$ are correlated through equation 
(\ref{s8w}), the observed cluster number constraints. Though the 
correlation in equation (\ref{s8w}) is confirmed by the statistics of 
cluster abundances, we see some breakup of this relation in lensing 
statistics. The breakup is most prominent for the NFW case.

\section{Comparison with Observations}
\label{section5}
To compare with observations we must consider the effect of magnification
bias \citep{tur84,koc95,bar01}. According to \citet{tur84}, the factor $B$
by which lensed objects at redshift $z_s$ will be overrepresented in any
particular observed sample may be written as
\begin{eqnarray}
    B = {\int_0^\infty S(f) \int_{A_m}^\infty A^{-1} P(A) N_{z_s}(f/A) 
        dA df \over \int_0^\infty S(f) N_{z_s}(f) df}\,,
    \label{mag_bias}
\end{eqnarray}
where $f$ is the observable flux of source objects, $S(f)$ is the 
selection function, $A_m$ is the minimum total flux amplification for a
multiple imaged source object, $P(A)$ is the probability density for a
greater amplification $A$, and $N_{z_s}(f) df$ is the number of source
objects in the sky at at redshift $z_s$ with unlensed flux lying between
$f$ and $f+df$. If the
sample of source objects has a flux distribution with a single power-law
$N_{z_s}\propto f^{-\beta}$,  the calculation of the magnification bias 
factor becomes extremely simple
\begin{eqnarray}
    B = \int_{A_m}^\infty A^{\beta-1} P(A) dA\,,
    \label{bias1}
\end{eqnarray}
which is independent of the selection function. Quite generally $P(A)$ is
given by $P(A) = 2 A_m^2 A^{-3}$ for $A\ge A_m$ \citep{tur84,sch92}. Then, 
if the sample of source objects has a power-law flux distribution, the 
magnification bias is simply
\begin{eqnarray}
    B = {2\over 3-\beta}\,A_m^{\beta -1}\,.
    \label{bias2}
\end{eqnarray}

For a SIS lens the total magnification is $A = 2/|y|$ so $A_m = 2$ since
multiple images are formed only if $y\le 1$ \citep{tur84,sch92}. For a NFW
or GNFW lens, the calculation of the total magnification is straightforward
but complex, and to determine its minimum is not easy. Here we estimate 
the minimum of the total magnification for a NFW/GNFW lens with
\begin{eqnarray}
    A_m \approx {2 x_0\over y_{\rm cr}}\,,
    \label{nfw_am}
\end{eqnarray}
where $x_0$ is the positive root of $y(x) = 0$, $y_{\rm cr} = -y(x_{\rm 
cr})$ and $x_{\rm cr}$ is the positive root of $dy/dx = 0$ (see subsection
\ref{section4.1} and Fig. \ref{fig4}). Note equation (\ref{nfw_am}) is
accurate for a SIS lens. Using equation (\ref{signfw}) and equation 
(\ref{deth}), equation (\ref{nfw_am}) can be written as
\begin{eqnarray}
    A_m \approx \Delta\theta\, D_L^A \left({\pi\over\sigma}\right)^{1/2}\,,
    \label{nfw_am2}
\end{eqnarray}
where $\sigma$ is the lensing cross-section.

For a SIS lens $A_m = 2$ which is a constant number, thus $B$ is 
independent of the redshift and mass of the lens and the observable lensing
probability $P_{\rm obs}$ is related to the intrinsic lensing probability 
$P$ simply by $P_{\rm obs} = B P$ \citep{tur84}. But for a NFW/GNFW lens, 
$A_m$ and $B$ depend on both the redshift and the mass of the lens, thus 
$P_{\rm obs}$ is related to $P$ by an integration
\begin{eqnarray}
    P_{\rm obs}(>\Delta\theta) = \int\int B\, {d^2 P(>\Delta\theta)\over 
       dM dz}\, dM dz\,,
    \label{pobs}
\end{eqnarray}
where the intrinsic lensing probability $P$ is given by equation (\ref{ip})
and equation (\ref{dpz}). The average magnification bias $\overline{B} 
\equiv P_{\rm obs}/ P$ is then a function of $\Delta\theta$.

Currently the largest uniformly selected sample of gravitational lensing
system is the Cosmic Lens All-Sky Survey (CLASS; Browne \& Myers 2000). 
The sample comprises 11685 flat-spectrum radio sources whose flux 
distribution is given by a power-law $N_{z_s}(f)\propto f^{-2.1}$ 
\citep{hel00,rus00}. The redshift distribution of the CLASS sample is
not known, but \citet{mar00} has reported redshifts for a small subsample 
of $42$ sources. They have found a mean source redshift of $\langle z_s
\rangle = 1.27$, which is comparable to that found in other radio surveys
at comparable fluxes. To date a total of $18$ multiple image gravitational 
lenses have been discovered in the combined JVAS\footnote{Jodrell-VLA 
Astrometric Survey, see \citet{kin99} and references therein.} and CLASS 
sample, all have image separations $\Delta\theta<3^{\prime\prime}$
\citep{bro00,hel00}. An explicit search for lenses with images separations 
$6^{\prime\prime}\le\Delta\theta\le 15^{\prime\prime}$ has found no lenses
\citep{phi00}. Among the $18$ discovered lens systems, one (B2114+022) is 
questionable \citep{hel00} so we exclude it from our analyses. The 
remaining $17$ lens systems are shown in Fig. \ref{fig11} as a histogram.

To compare our results with the CLASS survey, we assume all source objects
in the sample are at $z_s = 1.27$. For the CLASS sample $\beta \approx 
2.1$ \citep{rus00}, so from equation (\ref{bias2}) we have
\begin{eqnarray}
    B \approx 2.22\, A_m^{1.1}\,.
    \label{bias_c}
\end{eqnarray}
For SIS lenses $A_m = 2$, then we have $B\approx 4.76$. For NFW/GNFW 
lenses we calculate $A_m$ and $B$ using equation (\ref{nfw_am2}) and 
equation (\ref{bias_c}). Then we calculate the observable lensing 
probability using equation (\ref{pobs}). The results for SIS lenses and 
NFW lenses are shown in Fig. \ref{fig11} alternatively for SCDM with
$\Omega_m = 1$ and $\sigma_8 = 0.5$; OCDM with $\Omega_m = 0.3$,
$\Omega_\Lambda = 0$, and $\sigma_8 = 0.85$; LCDM with $\Omega_m = 0.3$,
$\Omega_\Lambda = 0.7$, and $\sigma_8 = 0.95$. Again, we assume the
Hubble constant $h=0.7$. The concentration parameters for the NFW lenses
are $c_1 = 5$ for SCDM, $9$ for OCDM, and $7$ for LCDM. For comparison,
we have also shown the result for a LWDM model with a
dashed-dotted curve, which has the same parameters as the LCDM model
except that for the LWDM we assume a power-law spectrum $P_k\propto
k^{-2}$ and take a smaller concentration parameter $c_1=3.5$. The null
results for the JVAS/CLASS survey for $6^{\prime\prime}\le\Delta\theta
\le 15^{\prime\prime}$ is shown with a horizontal straight line with a
downward arrow indicating that is an upper limit. Fig. \ref{fig11}
clearly shows that, all SIS models predict too many lenses for large 
splitting angles. While for large splitting angles NFW models are
consistent with the JVAS/CLASS results in the sense that the NFW
predictions are below the upper limit put by JVAS/CLASS, for small
splitting angles the NFW predictions are far below the observational
results of JVAS/CLASS. So, both SIS models and NFW models
cannot explain the observations. We have also tried GNFW models with
$1<\alpha<2$, neither of them can explain the observations on both
small splitting angles and large splitting angles. All the
models either predict too many lenses with large splitting angles, or
too few lenses with small splitting angles. Neither of them can produce
the observed large ratio of the number of small splittings to the number
of large splittings.

For the same cosmological and lens models we have also calculated the
average magnification bias $\overline{B} = P_{\rm obs}/P$, the results 
for NFW models and SIS models are shown in Fig. \ref{fig12}. We see that, 
the magnification bias for NFW lenses is bigger than the magnification 
bias for SIS lenses by about one order of magnitude. Though the 
magnification bias for SIS lenses is a constant for all cosmological 
models, the magnification bias for NFW lenses depends on cosmological 
models and slowly decreases as the splitting angle increases.

The above results strongly suggest that the halos in the real universe 
cannot be described by a single universal density profile. There must be 
at least two populations of ``halos'' in the universe: small mass halos 
with a steep inner density slope, and large mass halos with a shallow inner 
density slope. To test this conjecture, we have calculated the following 
additional lens model: lenses with mass $M<M_c$ have the SIS profile 
(which would produce the flat rotation curves seen in galaxies), but 
lenses with mass $M>M_c$ have the NFW
profile. The critical mass $M_c$ for transition from SIS to NFW is
determined by fitting our results with the JVAS/CLASS observations. Our 
fitting results indicate that $M_c\sim 10^{13}\, h^{-1}M_{\odot}$. 
Interestingly, this critical mass is very close to the cutoff mass of
halos below which cooling of the corresponding baryonic component will
lead to concentration of the baryons to the inner parts of the mass
profile \citep{ree77,blu86,por00}. The results for the case with
$M_c = 10^{13}\, h^{-1}M_{\odot}$ are shown in Fig. \ref{fig13}, which fit
the JVAS/CLASS observations reasonably well. The cosmological models are
the same as those in Fig. \ref{fig11}. The NFW halos (with $M > 10^{13}\,
h^{-1}M_{\odot}$) have the same concentration parameters as those in Fig.
\ref{fig11}. Though all the three models are consistent with the JVAS/CLASS
$6^{\prime\prime} - 15 ^{\prime\prime}$ survey, the SCDM model obviously
produces too many lenses with small splitting angles. It is interesting
that while SCDM produces the most small splittings, it produces the fewest
large splittings. We have calculated the expected total number of lenses 
with $\Delta\theta >0.3^{\prime\prime}$ for the JVAS/CLASS sample, the 
results are
\begin{eqnarray}
    N_{\rm lens}(>0.3^{\prime\prime}) = \left\{\begin{array}{ll}
       37, & \mbox{for SCDM}\\
       12, & \mbox{for OCDM}\\
       16, & \mbox{for LCDM}
       \end{array}
       \right.\,.
\end{eqnarray}
The expected number of lenses for the LCDM model is remarkably close to 
the number of lenses observed by the JVAS/CLASS survey which is $17$ if we
exclude the questionable lens of B2114+022 from the sample.

\section{Discussion and Conclusions}
\label{section6}
With the semi-analytical approach we have calculated the probability for 
forming multiple images of a remote source object gravitationally lensed 
by foreground dark halos. The mass density profile of a halo is alternatively 
taken to be described by the SIS profile, the NFW profile, and the GNFW 
profile. The mass function of halos is assumed to be given by the
Press-Schechter function. The cosmological model is
alternatively the SCDM model, the OCDM model, and the LCDM model. Our 
results show that the lensing probability is very sensitive to the density 
profile of lenses (dark halos). The more the mass distribution concentrates 
toward the center, the higher lensing probability the density profile gives 
rise, which is clearly seen in Fig. \ref{fig7} -- Fig. \ref{fig10}. 
Compared to SIS lenses, NFW lenses are extremely inefficient in producing
multiple images. For example, the lensing probability $P(>5^{\prime
\prime})$ for NFW halos is lower than the corresponding probability for
SIS halos by more than two orders of magnitudes. For GNFW halos, as 
$\alpha$ decreases, the differences among different cosmological models
become prominent (Fig. \ref{fig8}), especially the difference between the
SCDM model and the OCDM/LCDM model: when $\alpha = 1$ (i.e. for NFW halos)
the lensing probability for the SCDM model is lower than the lensing
probability for the OCDM/LCDM model by more than two orders of magnitudes.
This dramatic effect is dominantly caused by the fact that halos in a SCDM 
universe have the smallest concentration parameter \citep{bar98} and
for smaller $\alpha$ the lensing probability is more sensitive to the 
concentration parameter $c_1$: for the NFW case the lensing probability is 
extremely sensitive to $c_1$, especially for small $c_1$ (Fig. 
\ref{fig9}). Though not so dramatic as the dependence on the inner
slope, the lensing probability also shows dependence on the outer slope
of halos. Comparing the right panel of Fig. \ref{fig8} with Fig. \ref{fig3},
the lensing probability produced by GNFW halos with $\alpha = 2$ is 
somewhat higher than the lensing probability produced by SIS halos. This
mild difference is attributed to the different slopes of the $\alpha = 2$,
GNFW profile and the SIS profile at large radii (the $\alpha = 2$, GNFW 
profile and the SIS profile have the same slope at small radii). Similar
conclusions are obtained by \citet{wyi00} and \citet{kee01}. \citet{wyi00}
find that the optical depth to multiple imaging is a very sensitive 
function of the profile parameters and the GNFW profile exhibits 
degeneracies between profile parameters with respect to lensing statistics.
\citet{kee01} demonstrate that the lensing probability is determined almost 
entirely by the fraction of the halo mass that is contained within a 
fiducial radius that is $\sim 4\%$ of the virial radius.

\begin{figure}
\epsscale{0.82}
\plotone{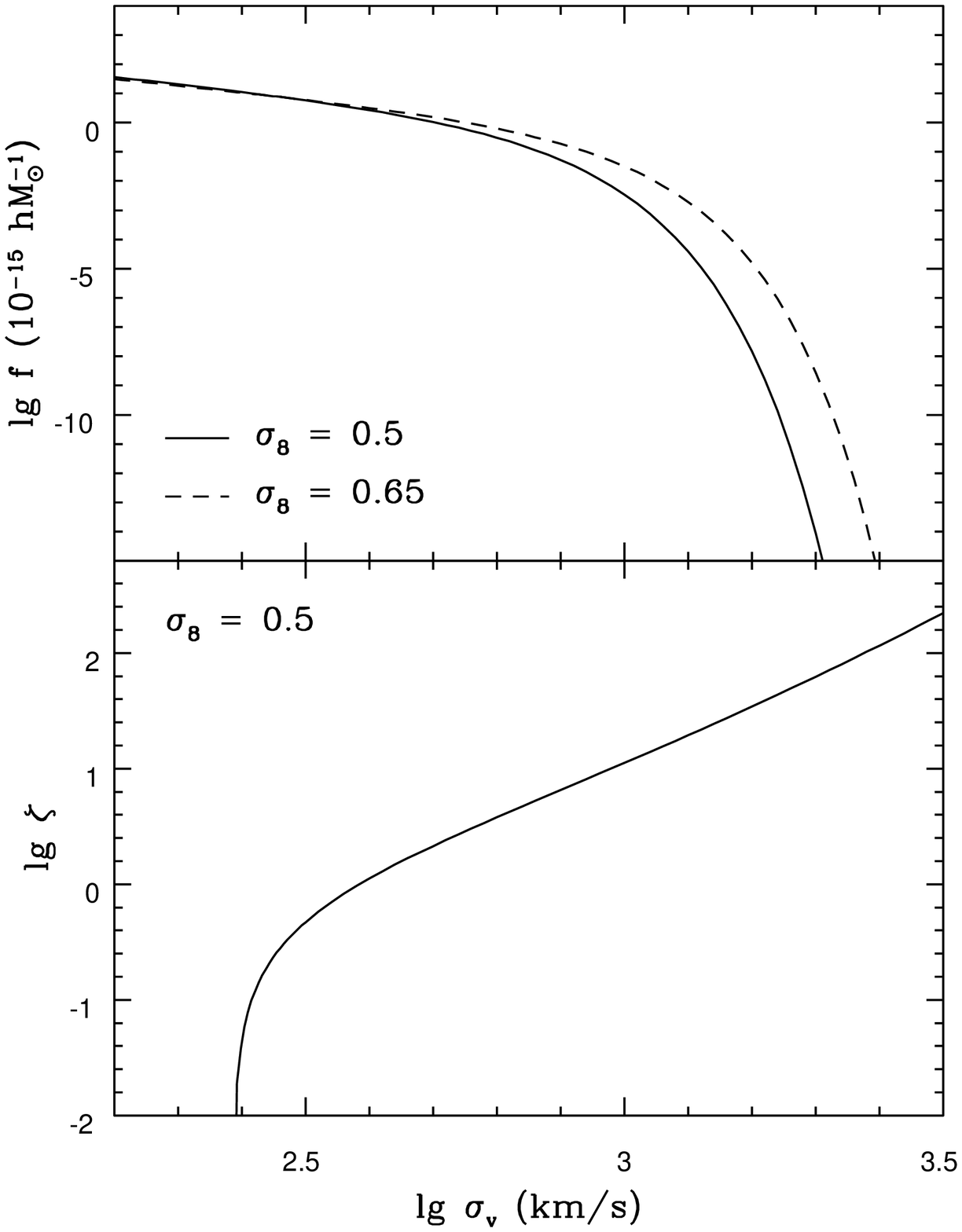}
\caption{Upper panel: The Press-Schechter function $f$ against the 
velocity dispersion $\sigma_v$ of dark halos at $z=0$ in a SCDM cosmology. 
The solid curve is for $\sigma_8 = 0.5$, the dashed curve is for $\sigma_8 
= 0.6$. Lower panel: $\zeta\equiv{\delta f/f\over\delta\sigma_8/\sigma_8}$  
as a function of $\sigma_v$, for a SCDM cosmology with $z=0$ and $\sigma_8 
= 0.5$. Both panels show that for large $\sigma_v$, a poor knowledge in 
$f$ may give a good estimation of $\sigma_8$. Since lensing is produced by
the high $\sigma_v$ part of the distribution, the number of lenses 
sensitively constrains $\sigma_8$.
\label{fig1}}
\end{figure}

\begin{figure}
\epsscale{1}
\plotone{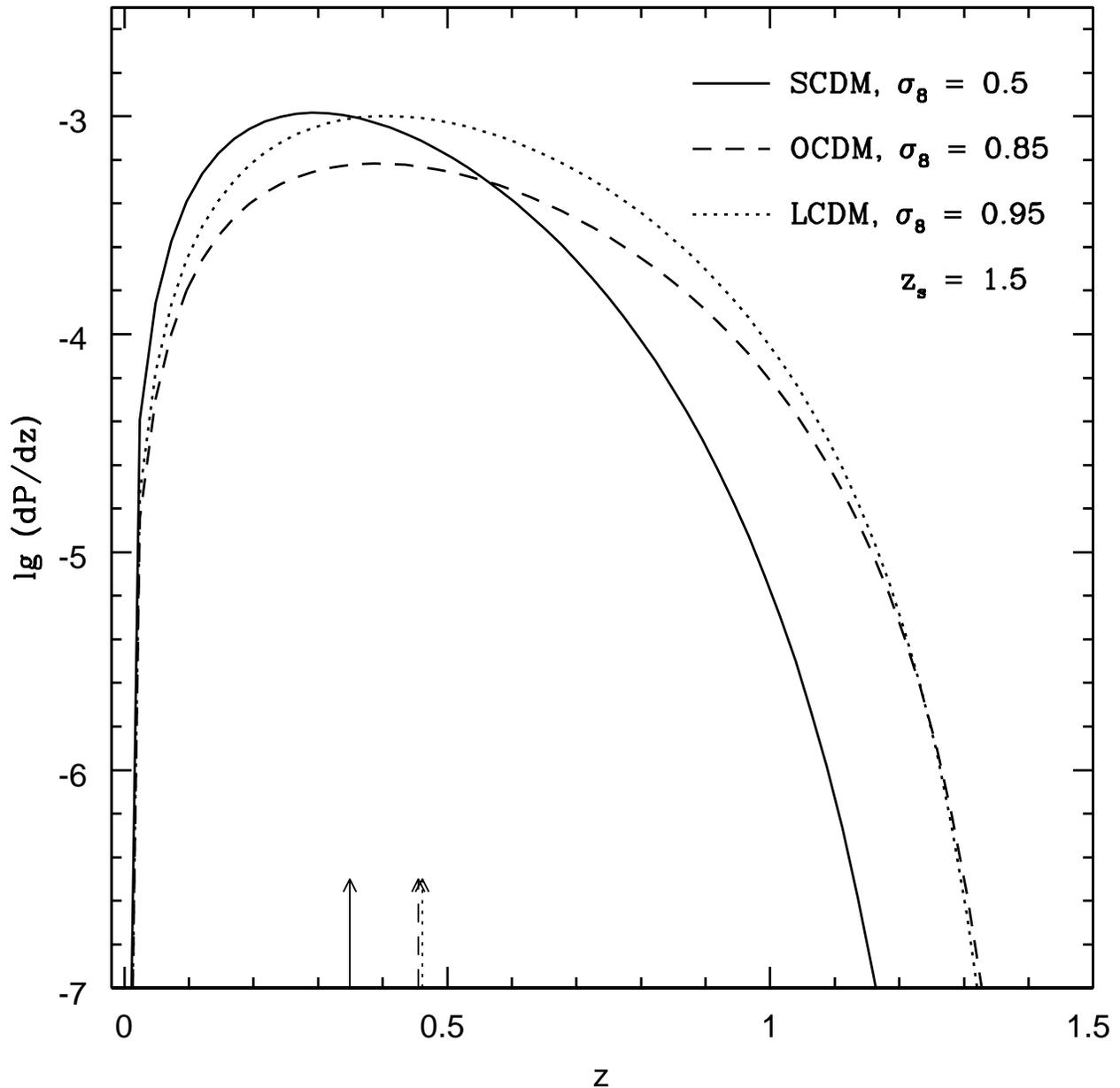}
\caption{The differential lensing probability $dP/dz$ for $\Delta\theta>
5^{\prime\prime}$ as a function of lenses' redshift $z$. The source object
is assumed to be at $z_s = 1.5$. The lens objects are SIS halos. The 
cosmological models are alternatively: SCDM with $\Omega_m = 1$ and 
$\sigma_8 = 0.5$ (solid line); OCDM with $\Omega_m = 0.3$, $\Omega_\lambda 
= 0$, and $\sigma_8 =0.85$ (dashed line); LCDM with $\Omega_m = 0.3$, 
$\Omega_\lambda = 0.7$, and $\sigma_8 = 0.95$ (dotted line). The Hubble 
constant is $h=0.7$. As expected, the probable position of the lens is at
significantly higher redshifts for the low $\Omega_m$ models than for SCDM 
-- note arrows which show median expected redshifts.
\label{fig2}}
\end{figure}

\begin{figure}
\epsscale{1}
\plotone{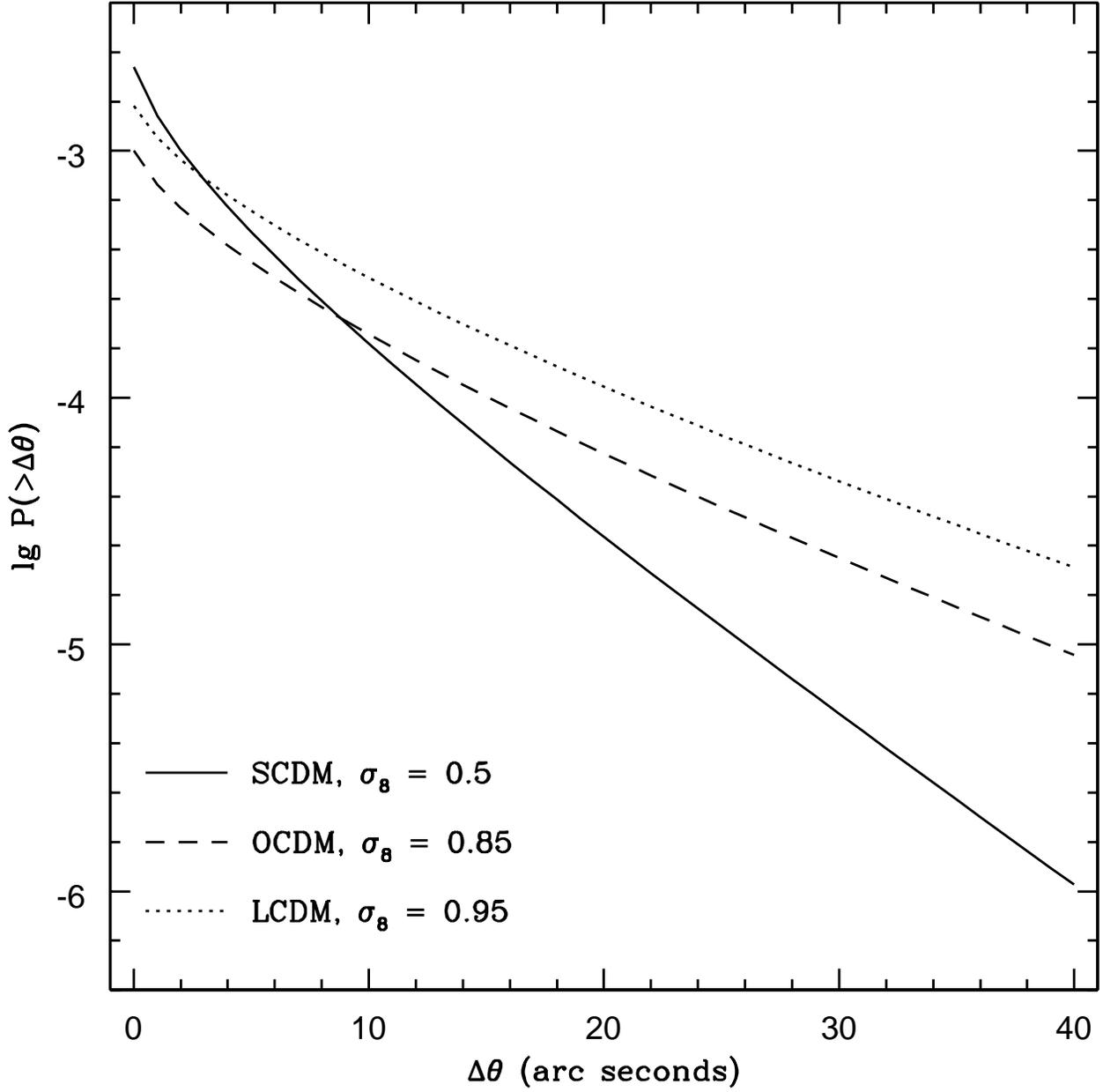}
\caption{The integral lensing probability $P(>\Delta\theta)$ as a function 
of $\Delta\theta$. The models are the same as those in Fig. \ref{fig2}.
\label{fig3}}
\end{figure}

\begin{figure}
\epsscale{0.98}
\plotone{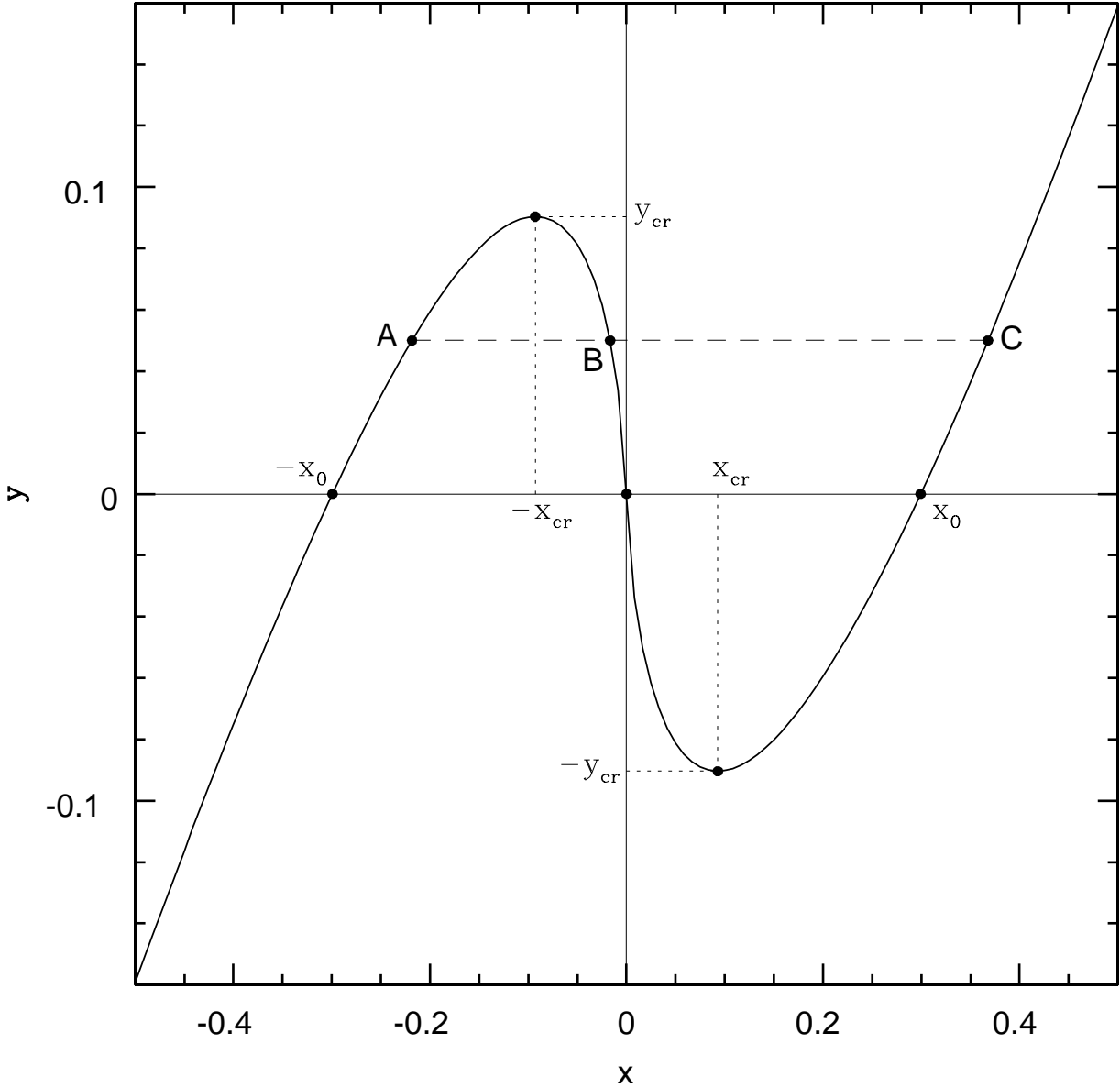}
\caption{The lensing equation for the GNFW profile, i.e. equation 
(\ref{le1}) in the text. Here shows the case with $\alpha = 1.2$ and
$\mu_s = 1$. The horizontal axis is $x$, which labels the position in the 
lens plane; the vertical axis is $y$, which labels the position in the
source plane. The points $(x_{\rm cr},-y_{\rm cr})$ and $(-x_{\rm cr},
y_{\rm cr})$ satisfy $dy/dx = 0$. The non-zero roots of $y(x) = 0$ are 
$\pm x_0$. Three images are formed when $|y|<y_{\rm cr}$, two images are 
formed when $|y|=y_{\rm cr}$, one image is formed when $|y|>y_{\rm cr}$. 
So, multiple images are formed when $|y|\le y_{\rm cr}$, an example is 
shown with the horizontal long dashed line ABC -- which has three images: 
A, B, and C. In the paper we consider the splitting angle between the two 
outside images, i.e the splitting angle between A and C.
\label{fig4}}
\end{figure}

\begin{figure}
\epsscale{1}
\plotone{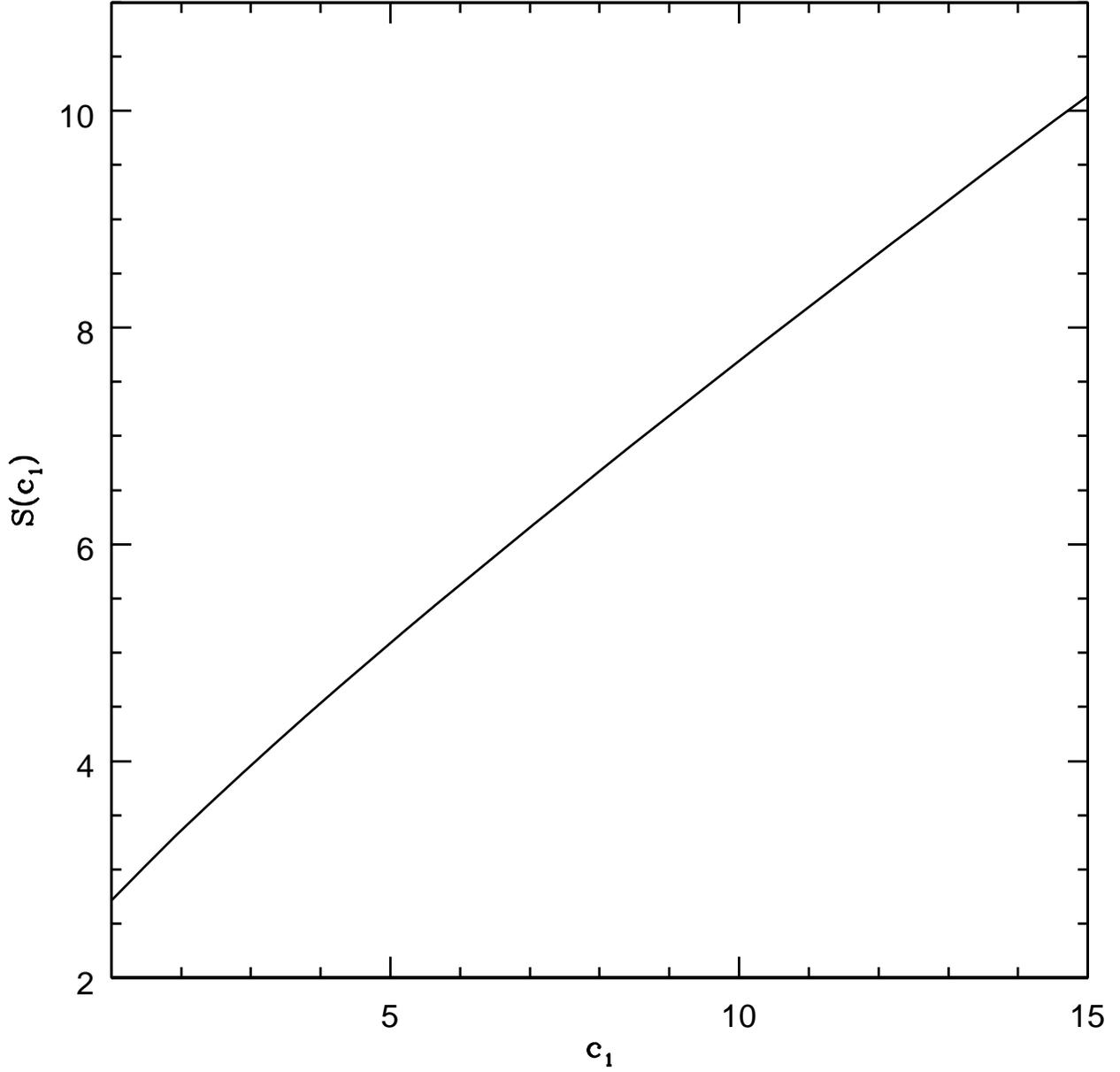}
\caption{$S(c_1) \equiv 1/\eta(c_1)^2$, where $c_1$ is the concentration
parameter, $\eta(c_1)\equiv r_{1/2}/r_{200}$ is determined by equation
(\ref{rel}), where $r_{1/2}$ is the half-mass radius of a cluster. 
$S(c_1)$ is proportional to the surface mass density of a halo at the 
half-mass radius (eq. [\ref{shalf}]).
\label{fig5}}
\end{figure}

\begin{figure}
\epsscale{0.98}
\plotone{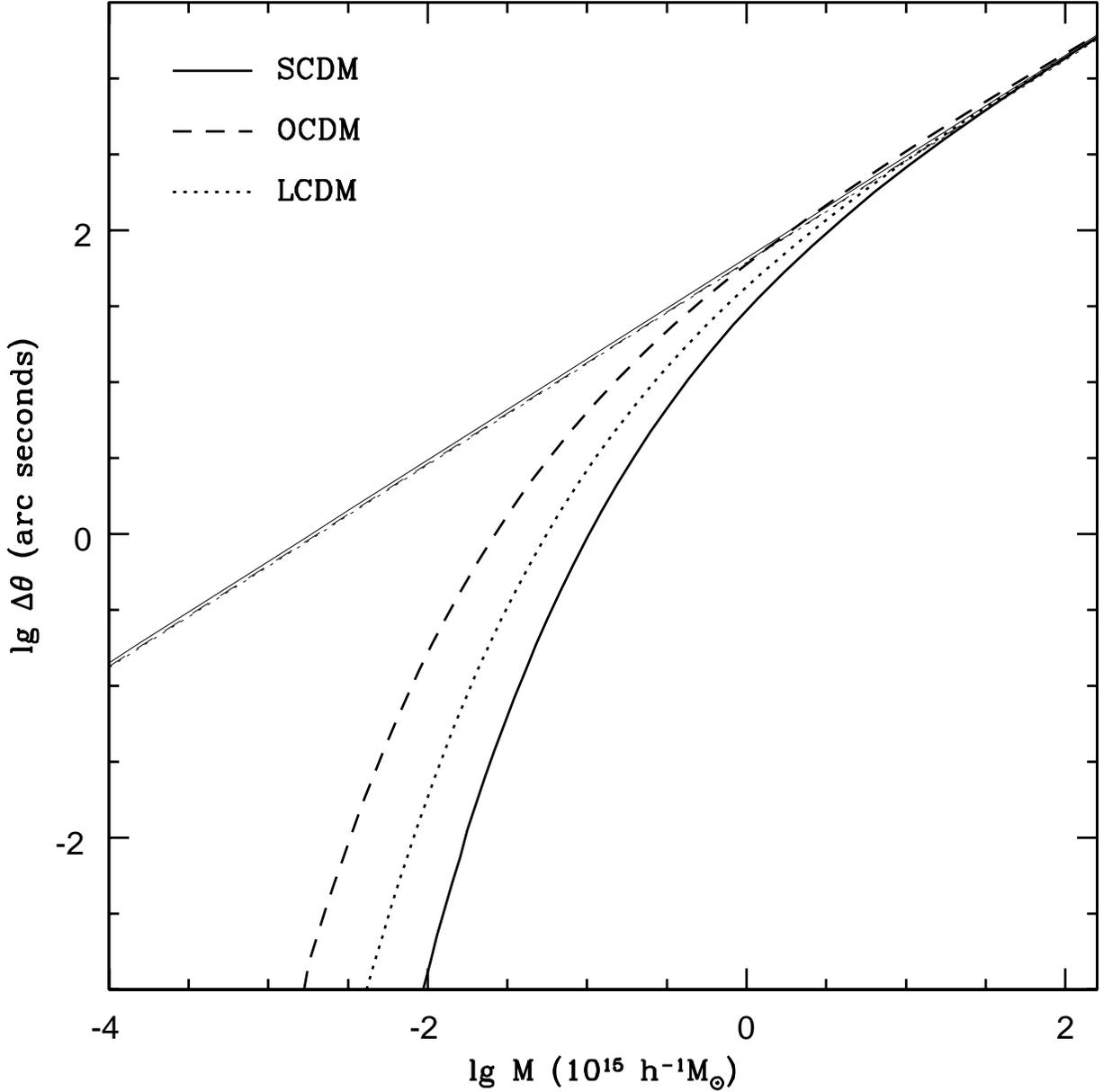}
\caption{Splitting angle $\Delta\theta$ as a function of $M$ -- the mass 
of the lens object. The thick lines show the splitting angle produced by
a NFW lens, the thin lines show the splitting angle produced by a SIS 
lens. The source object is at $z=1.5$, the lens object is at $z=0.3$. The 
cosmological models are alternatively: SCDM with $\Omega_m=1$ (solid 
lines), OCDM with $\Omega_m=0.3$ and $\Omega_{\Lambda}=0$ (dashed lines), 
LCDM with $\Omega_m=0.3$ and $\Omega_{\Lambda}=0.7$ (dotted lines). For 
the NFW case, the concentration parameters are alternatively $5$ for SCDM,
$9$ for OCDM, and $7$ for LCDM. Notice the enormous sensitivity at small
splittings and small (e.g. galactic) mass halos to the steepness of the
inner part of the mass profile.
\label{fig6}}
\end{figure}

\begin{figure}
\epsscale{1.0}
\plotone{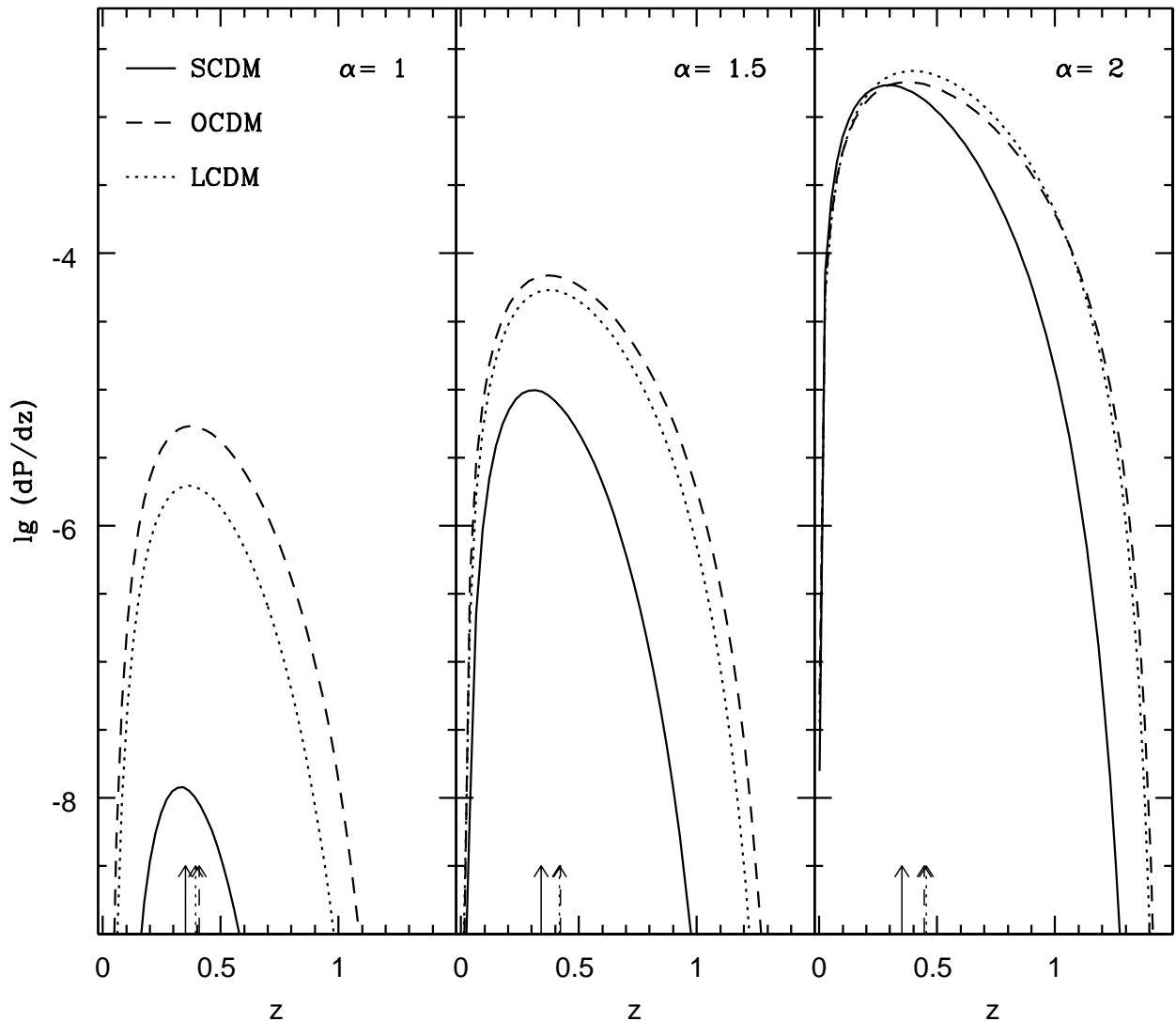}
\caption{The differential lensing probability $dP/dz$ for $\Delta\theta>
5^{\prime\prime}$ as a function of lenses' redshift $z$. The source object
is at $z_s = 1.5$. The lens objects are alternatively NFW halos (left
panel, i.e. GNFW halos with $\alpha=1$), GNFW halos with $\alpha = 1.5$
(central panel), and GNFW halos with $\alpha = 2$ (right panel). The
concentration parameters are alternatively given by equation 
(\ref{con_nfw}), equation (\ref{con_nfw2}), and equation (\ref{con_nfw3}). 
The cosmological models are alternatively: SCDM with $\Omega_m =1$ and
$\sigma_8 = 0.5$; OCDM with $\Omega_m = 0.3$, $\Omega_\Lambda = 0$, and
$\sigma_8 = 0.85$; OCDM with $\Omega_m = 0.3$, $\Omega_\Lambda = 0.7$, 
and $\sigma_8 = 0.95$. The Hubble constant is $h=0.7$. The extreme 
sensitivity of lensing to the slope of the inner profile is evident.
Arrows show median expected redshifts.
\label{fig7}}
\end{figure}

\begin{figure}
\epsscale{1.0}
\plotone{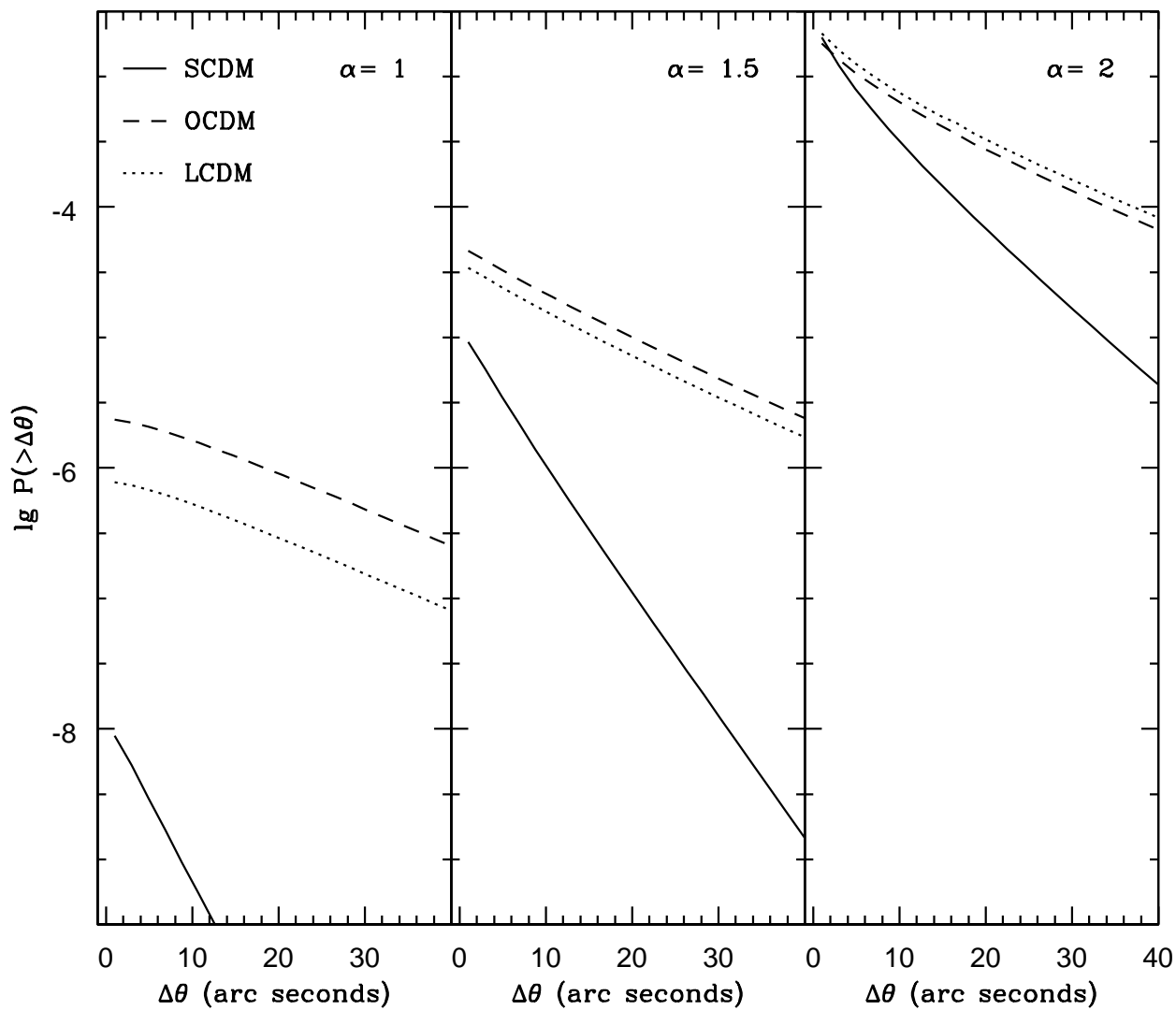}
\caption{The integral lensing probability $P(>\Delta\theta)$ as a function 
of $\Delta\theta$. The models are the same as those in Fig. \ref{fig7}.
Note the relatively slow fall-off of probability with increasing splitting
angle for the OCDM and LCDM cosmologies: for all $\alpha$ the number of
$10^{\prime\prime}$ splittings expected is not very much smaller than the 
number of $1^{\prime\prime}$ splittings expected.
\label{fig8}}
\end{figure}

\begin{figure}
\epsscale{1.0}
\plotone{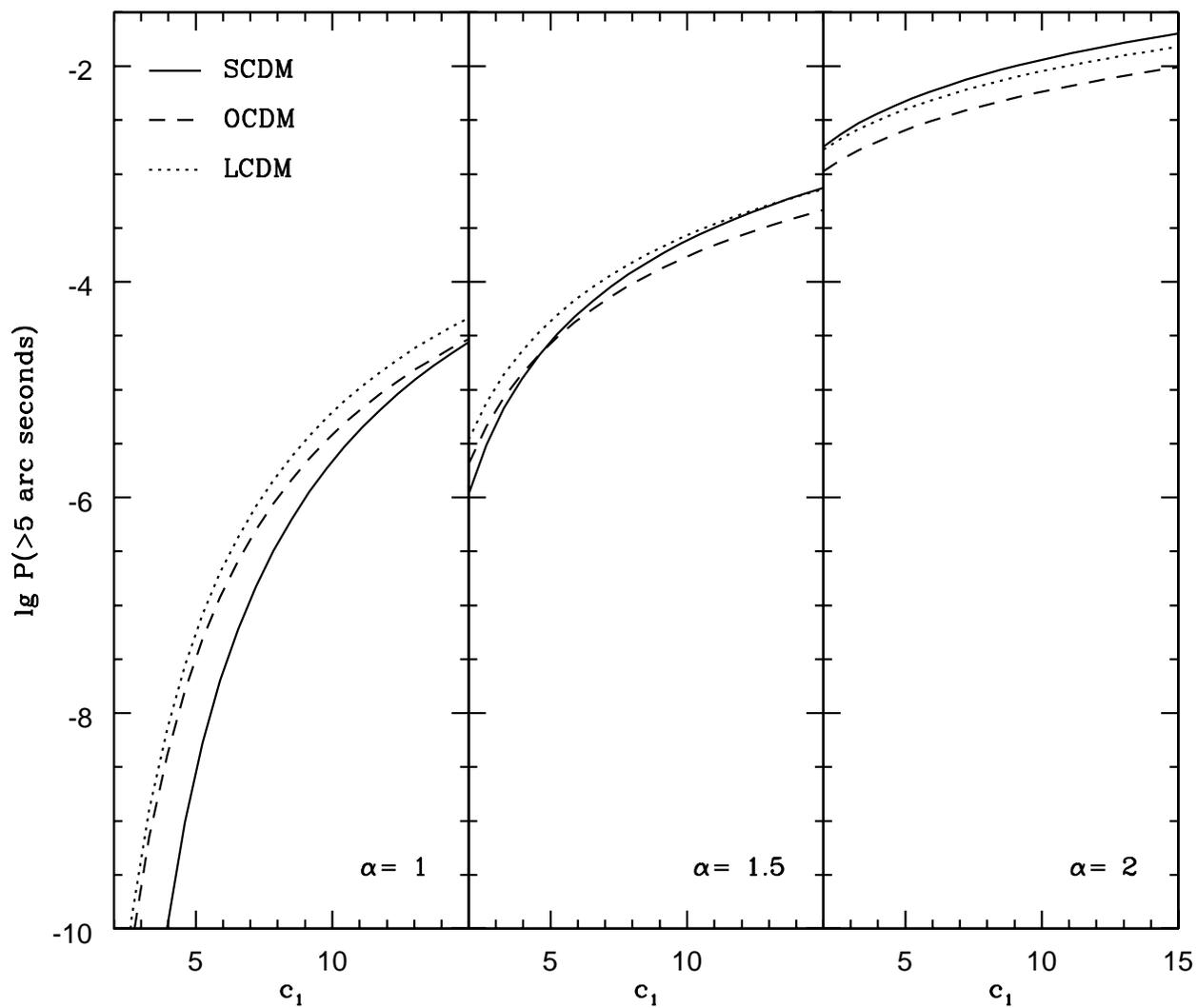}
\caption{The integral lensing probability $P(\Delta\theta>
5^{\prime\prime})$ as a function of lenses' concentration parameter $c_1$.
The cosmological models are the same as those in Fig. \ref{fig7}. The 
source object is at $z_s=1.5$. The lens objects are alternatively NFW 
halos (left panel), GNFW halos with $\alpha = 1.5$ (central panel), and 
GNFW halos with $\alpha = 2$ (right panel).
\label{fig9}}
\end{figure}

\begin{figure}
\epsscale{1.0}
\plotone{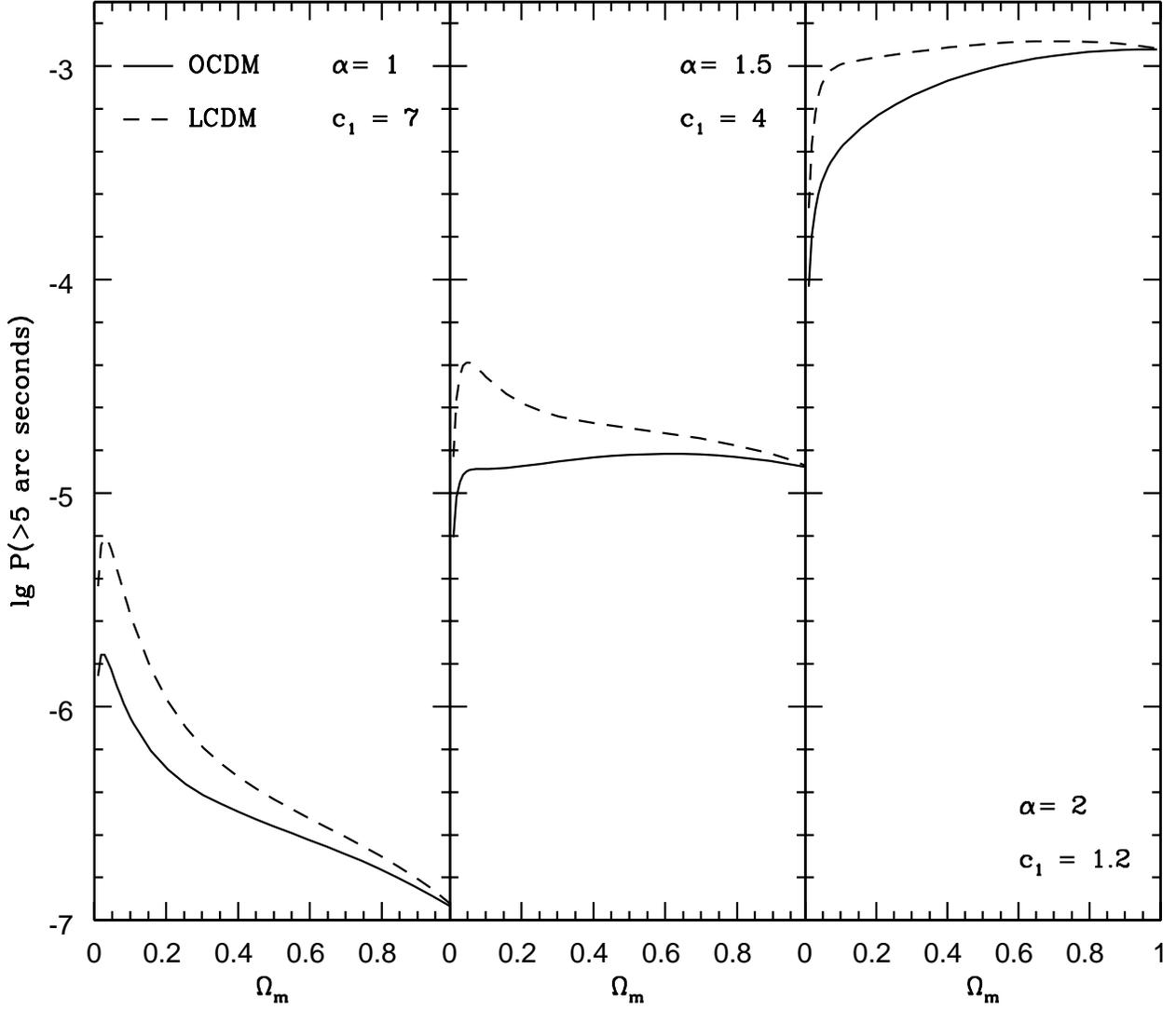}
\caption{The integral lensing probability $P(\Delta\theta>
5^{\prime\prime}$) as a function of $\Omega_m$ when $\Omega_m$ and 
$\sigma_8$ are constrained by equation (\ref{s8w}). The lens models are
the same as those in Fig. \ref{fig7}. The source object is at $z_s=1.5$.
The cosmological models are OCDM (solid lines) and LCDM (dashed lines).
The Hubble constant is $h=0.7$. Note that, for models normalized to
give the correct $z=0$ cluster abundances, the sensitivity of the 
lensing probability to the halo profile shape is far stronger than
to the matter density $\Omega_m$.
\label{fig10}}
\end{figure}

\begin{figure}
\epsscale{0.9}
\plotone{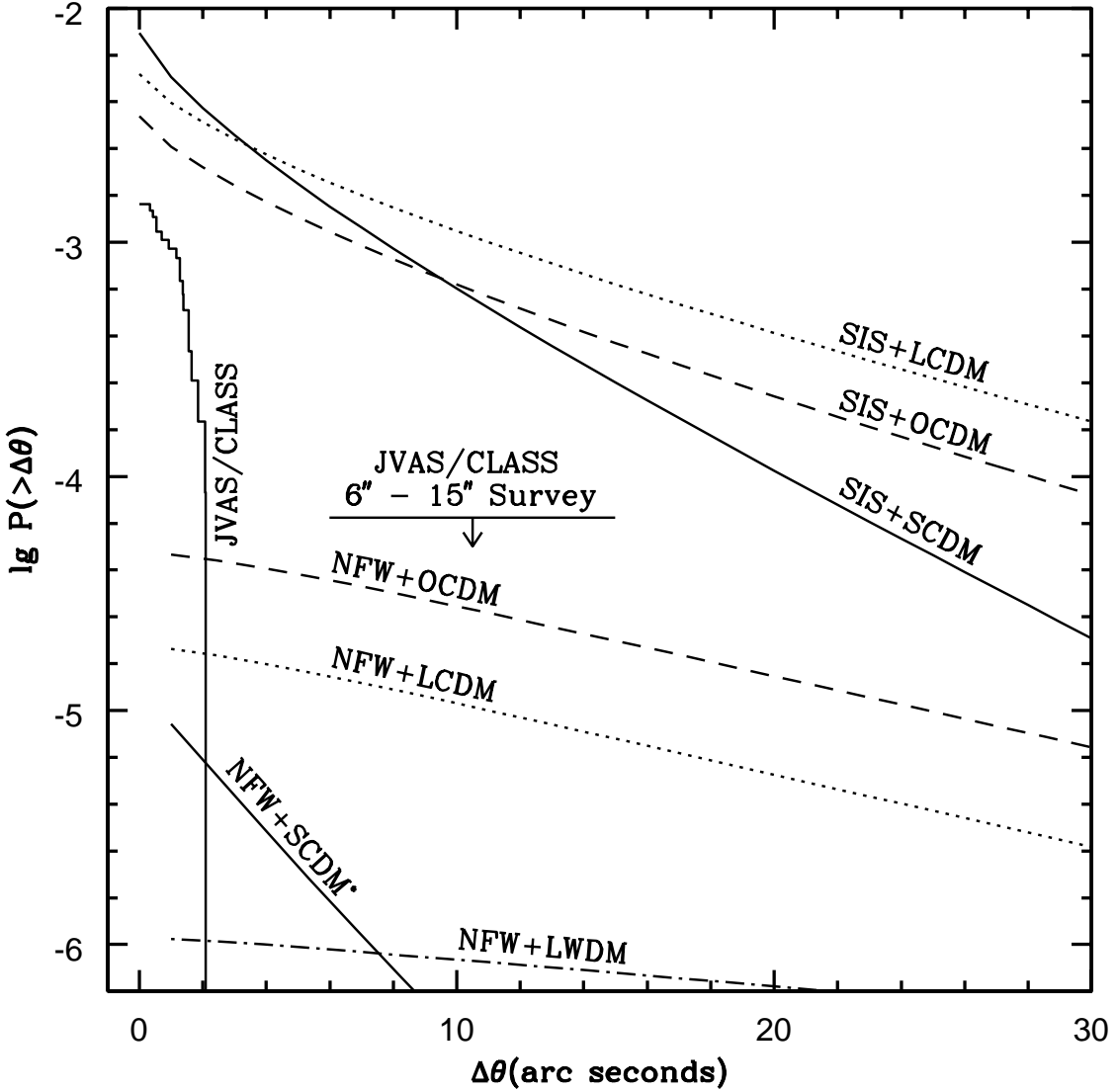}
\caption{Comparison of our semi-analytical results with the JVAS/CLASS 
survey. The JVAS/CLASS results are shown with the histogram. The null 
result for
lenses with $6^{\prime\prime}\le\Delta\theta\le 15^{\prime\prime}$ is
shown with the horizontal line with a downward arrow indicating that is 
an upper limit. Our semi-analytical results (allowing for amplification
bias) are shown with solid curves
(SCDM with $\Omega_m = 1$ and $\sigma_8 = 0.5$), dashed curves (OCDM with 
$\Omega_m = 0.3$, $\Omega_\Lambda = 0$, and $\sigma_8 = 0.85$), and dotted 
curves (LCDM with $\Omega_m = 0.3$, $\Omega_\Lambda = 0.7$, and $\sigma_8 
= 0.95$). Two groups of lenses models are shown: SIS lenses, and NFW 
lenses with the concentration parameter $c_1 = 5$ for SCDM, $9$ for OCDM, 
$7$ for LCDM. The NFW+SCDM lensing probability has been multiplied by a 
factor $10^{1.5}$ to fit it on the same scale as other models. For 
comparison we have also shown the result for a NFW+LWDM model with the
dashed-dotted curve, which has the same parameters as the NFW+LCDM model
except that $c_1=3.5$.
\label{fig11}}
\end{figure}

\begin{figure}
\epsscale{1.0}
\plotone{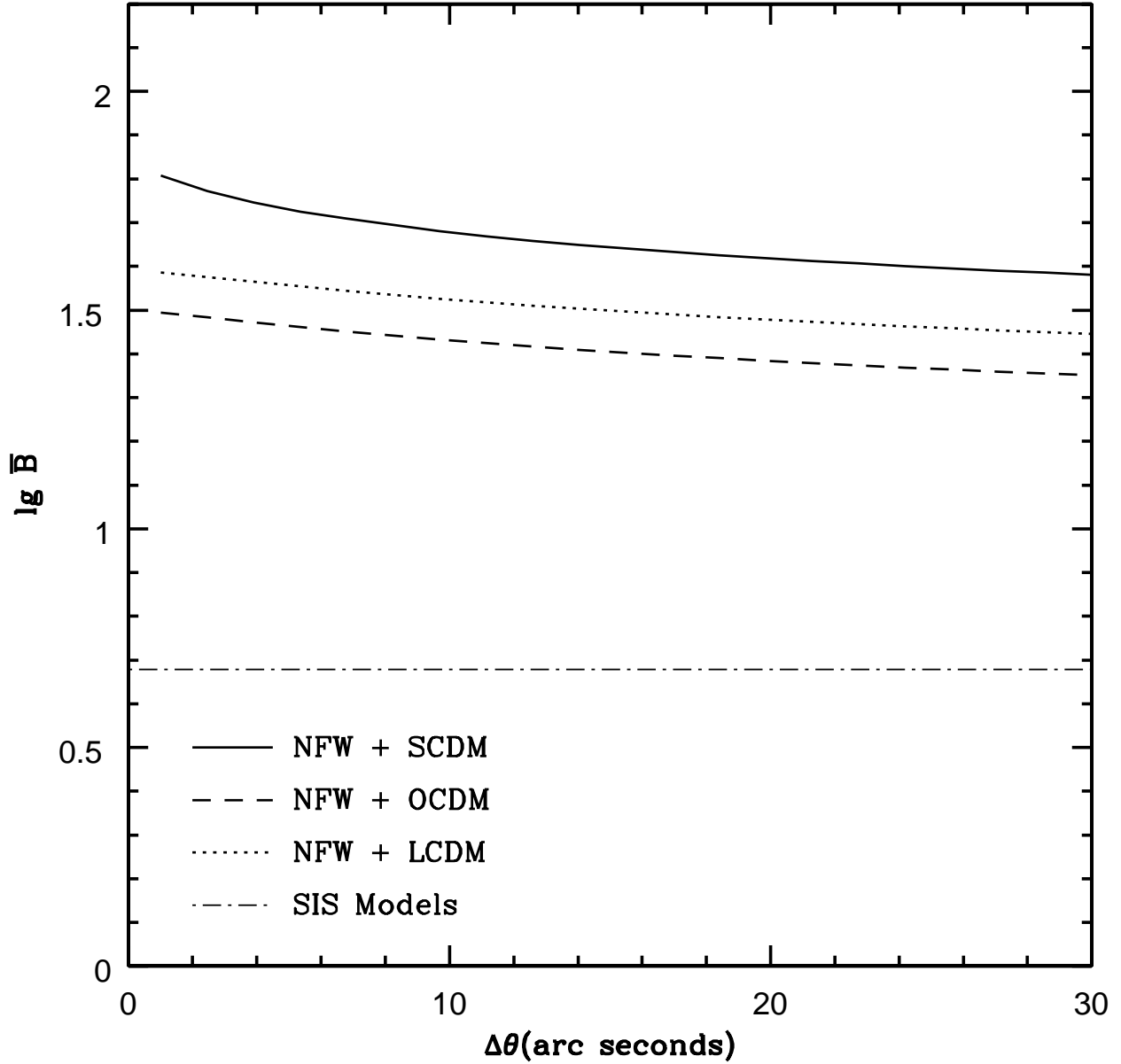}
\caption{The average magnification bias as a function of splitting angles.
The solid, dashed, and dotted lines are for NFW lenses, the thin
dashed-dotted line is for SIS lenses. The cosmological and lens models
are the same as those in Fig. \ref{fig11}. The magnification bias for SIS 
lenses is a constant independent of cosmological models.
\label{fig12}}
\end{figure}

\begin{figure}
\epsscale{1}
\plotone{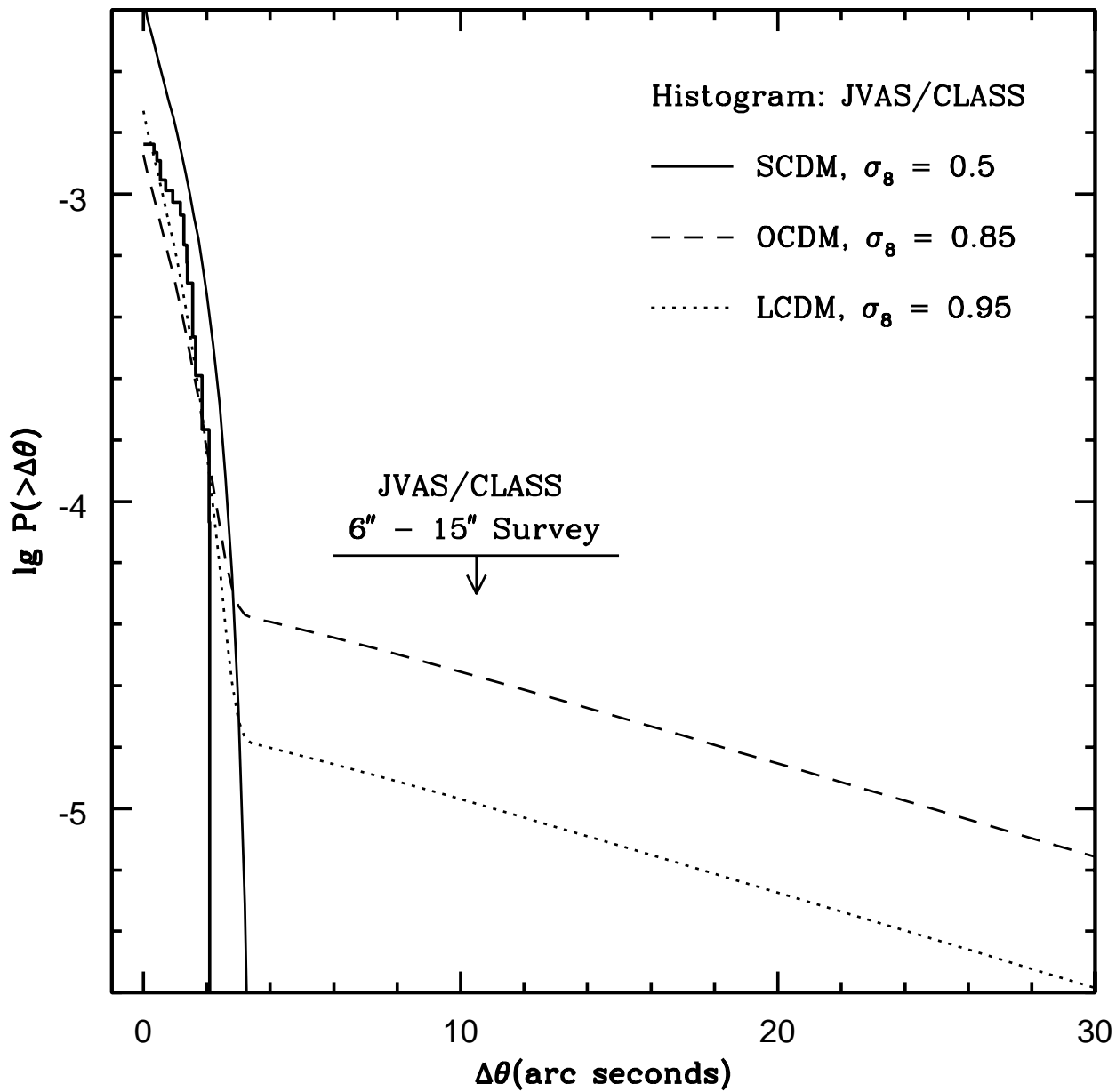}
\caption{Fitting of the two-population models to the JVAS/CLASS results. 
The cosmological models are the same as those in Fig. \ref{fig11}. The 
lens models are: SIS halos with $M< 10^{13}\, h^{-1}M_{\odot}$, NFW halos 
with $M> 10^{13}\, h^{-1}M_{\odot}$. The concentration parameters for NFW
halos are alternatively $c_1=5$ for SCDM, $9$ for OCDM, and $7$ for LCDM.
\label{fig13}}
\end{figure}

Our results also show the dependence of the lensing probability on the
cosmological parameters, which is particularly manifested in the SIS case.
In Fig. \ref{fig3}, for small splitting angles the lensing probability in
the SCDM universe is higher than the lensing probability in the OCDM/LCDM
universe, while for large splitting angles the lensing probability in the
SCDM universe is much lower than the lensing probability in the OCDM/LCDM
universe. This is explained by the fact that the Press-Schechter function
is sensitive to $\sigma_8$ only for large mass halos (Fig. \ref{fig1}).
Since small splitting angles are produced by small mass halos, and for small
mass halos the Press-Schechter function is not sensitive to $\sigma_8$, the
lensing probability is dominantly determined by the cosmic mean mass 
density $\Omega_m$; thus for small splitting angles the lensing probability
is highest for SCDM since the SCDM universe has the highest mean mass 
density. Since large splitting angles are produced by large mass halos and 
for large mass halos the Press-Schechter function is (exponentially) 
sensitive to $\sigma_8$, the lensing probability is dominantly determined 
by $\sigma_8$, thus for large splitting angles the lensing probability is 
lowest for SCDM since the SCDM universe has the smallest $\sigma_8$. For 
the NFW case, in Fig. \ref{fig6} we see that to produce the same small 
splitting angle the required
NFW halo mass is significantly larger than the required SIS halo mass. 
Thus, for the NFW case $\sigma_8$ takes effect for all $\Delta\theta\ge
1^{\prime\prime}$, which together with the concentration parameter $c_1$
overtakes the effect of $\Omega_m$ (see Fig. \ref{fig8}). The studies of
cluster abundances proposed a correlation between $\Omega_m$ and $\sigma_8$
as given by equation (\ref{s8w}). We have also tested the sensitivity of
the lensing probability to $\Omega_m$ when equation (\ref{s8w}) is 
satisfied. We see that in general the lensing probability varies with
$\Omega_m$; and for the NFW case the lensing probability is very sensitive 
to $\Omega_m$ (Fig. \ref{fig10}).

Our numerical results are summarized in Table \ref{tab1}, where we have not
included the effect of magnification bias (i.e. $P$ is the intrinsic
probability), the source object is assumed to be at $z_s=1.5$, the Hubble
constant is assumed to be $h=0.7$. To show the 
sensitivity of the lensing probability to the parameters $\sigma_8$, 
$\Omega_m$, $\alpha$, and $c_1$, we have evaluated the differentiation 
of lensing probability $P(>\Delta\theta)$ with respect to $\sigma_8$, 
$\Omega_m$, $\alpha$, and $c_1$ for the LCDM model at $\sigma_8 = 0.95$, 
$\Omega_m = 0.3$, $\Omega_{\Lambda} = 0.7$, $\alpha = 1.5$, and 
$c_1(z=0) = 4$ for $\Delta\theta =5^{\prime\prime}$ and $\Delta\theta=
10^{\prime\prime}$ respectively. The results are
\begin{eqnarray}
    {\delta P\over P} &=& 4.9 \left({\delta \sigma_8\over \sigma_8}
        \right)_{\Omega_m,\alpha,c_1} + 1.7\, \left({\delta \Omega_m\over 
        \Omega_m}\right)_{\sigma_8,\alpha,c_1} +11 \left({\delta \alpha
        \over \alpha}\right)_{\sigma_8,\Omega_m,c_1}\nonumber\\
        &&+3.1\left({\delta c_1\over c_1}\right)_{\sigma_8,
        \Omega_m,\alpha}\,,
        \label{diff1}
\end{eqnarray}
for $\Delta\theta=5^{\prime\prime}$, and
\begin{eqnarray}
    {\delta P\over P} &=& 5.8 \left({\delta \sigma_8\over \sigma_8}
        \right)_{\Omega_m,\alpha,c_1} + 1.8\, \left({\delta \Omega_m\over 
        \Omega_m}\right)_{\sigma_8,\alpha,c_1} +10 \left({\delta \alpha
        \over \alpha}\right)_{\sigma_8,\Omega_m,c_1}\nonumber\\
        &&+3.4\left({\delta c_1\over c_1}\right)_{\sigma_8,\Omega_m,
        \alpha}\,,
        \label{diff2}
\end{eqnarray}
for $\Delta\theta=10^{\prime\prime}$. Note the extreme sensitivity of the 
lensing probability to the slope of the inner profile: $(d\ln P/ d\ln 
\alpha)_{\sigma_8,\Omega_m, c_1} \approx 10$. Note also that, if $\Omega_m$
and $\sigma_8$ enter the lensing probability as a combination given by
equation (\ref{s8w}), then we should expect $(d\ln \sigma_8/ d\ln
\Omega_m)_{P,\alpha, c_1} \approx \gamma \approx 0.53$, but equation
(\ref{diff1}) and equation (\ref{diff2}) show rather smaller values of
$(d\ln \sigma_8/ d\ln\Omega_m)_{P,\alpha, c_1}$ ($0.34$ and $0.31$ 
respectively). Thus, in lensing statistics the degeneracy between
$\Omega_m$ and $\sigma_8$ in equation (\ref{s8w}) is broken. This is 
important since it allows us to separately determine the values of 
$\Omega_m$ and $\sigma_8$ in principle.

With the magnification bias being considered, none of the simple models can 
completely explain the JVAS/CLASS observations. The SIS models produce too 
many large splitting lenses, but the JVAS/CLASS observations have detected 
no lenses with $\Delta\theta > 3^{\prime\prime}$ \citep{hel00,phi00}. The 
null result for detecting large splitting lenses are true not only for 
JVAS/CLASS \citep{phi00}, but also for ARCS (Arcminute Radio Cluster-lens 
Search) which is aimed at looking for gravitational lensing events with 
images separation between $15^{\prime\prime}$ and $60^{\prime\prime}$
\citep{phi00a}.
While the NFW models produce very rare large splitting lenses which is
consistent with observations, they produce too few small splitting lenses
which is against observations since in the JVAS/CLASS survey at least $17$
lenses with $0.3^{\prime\prime}<\Delta\theta<3^{\prime\prime}$ have been
discovered \citep{hel00}. None of the models can explain the observable 
large ratio of the number of
small splittings to the number of large splittings. This strongly
suggests that there are at least two populations of halos in the universe:
small mass halos with a steep inner density slope, and large mass halos with
a shallow inner slope. We have constructed a very simple two-population 
halo model to test this conjecture: the mass density of halos with mass $<
10^{13}\,h^{-1} M_{\odot}$ is given by the SIS profile, while the mass 
density of halos with mass $> 10^{13}\,h^{-1}M_{\odot}$ is given by the
NFW profile. 
We find that the results for this model are reasonably consistent with the
JVAS/CLASS observations. In particular, the number of lenses with $\Delta
\theta>0.3^{\prime\prime}$ predicted by the LCDM model is $\approx 16$, 
which is remarkably close to the number of lenses observed by the JVAS/CLASS 
survey which is $17$. The SIS model predicts too many lenses ($\approx 37$), 
while the OCDM model predicts somewhat too few lenses ($\approx 12$). A 
similar compound model has also been considered by \citet{por00} to explain 
the results from CASTLE (CfA-Arizona Space Telescope Lens) survey.

In summary, with the semi-analytical approach we have shown that the
gravitational lensing probability is very sensitive to the mass density 
profile of lenses (especially in the central region), the mean mass density 
in the universe, and the amplitude of primordial fluctuations. Compared 
with the observation results of JVAS/CLASS, our calculations indicate that 
the halos in the real universe cannot be described by a single universal
universal density profile, there are at least two populations of halos in 
the universe: small mass halos with a steep inner density slope and large 
mass halos with a shallow inner density slope. Ultimately, of course, very
accurate address of the question is left open by this study.

\acknowledgments

We are grateful to E. L. Turner for many helpful discussions and comments,
to C. R. Keeton and P. Madau for helpful communications. We would like to
thank the anonymous referee whose comments led to the improvement of this 
work. This research was supported by the NSF  grants ASC-9740300 (subaward 
766) and AST-9803137.

\clearpage

\clearpage

\clearpage

\clearpage

\clearpage

\clearpage

\clearpage

\clearpage

\clearpage

\clearpage

\clearpage

\clearpage

\clearpage

\clearpage
\begin{deluxetable}{lccccc}
\tablewidth{0pt}
\tablecaption{Summary of the Results for the Lensing Probability \label{tab1}}
\tablehead{
\colhead{Models} & \colhead{$\lg P(2^{\prime\prime})$} & \colhead{$\lg 
	P(5^{\prime\prime})$} &
\colhead{$\lg P(10^{\prime\prime})$} & \colhead{$\lg P(20^{\prime\prime})$} &
\colhead{$\lg P(40^{\prime\prime})$}
}
\startdata
SCDM\tablenotemark{a} ($\alpha=1$) & $-$8.16 & $-$8.54 & $-$9.17 & $-$10.4 
	& $-$12.8\\
SCDM\tablenotemark{a} ($\alpha=1.5$) & $-$5.14 & $-$5.46 & $-$5.98 & $-$6.96 
	& $-$8.84\\
SCDM\tablenotemark{a} ($\alpha=2$) & $-$2.82 & $-$3.10 & $-$3.49 & $-$4.17 
	& $-$5.36\\
SCDM\tablenotemark{a} (SIS) & $-$3.00 & $-$3.33 & $-$3.78 & $-$4.56 
	& $-$5.97\\
OCDM\tablenotemark{b} ($\alpha=1$) & $-$5.64 & $-$5.68 & $-$5.79 & $-$6.04 
	& $-$6.60\\
OCDM\tablenotemark{b} ($\alpha=1.5$) & $-$4.37 & $-$4.49 & $-$4.66 & $-$5.00 
	& $-$5.62\\
OCDM\tablenotemark{b} ($\alpha=2$) & $-$2.81 & $-$2.98 & $-$3.20 & $-$3.56 
	& $-$4.17\\
OCDM\tablenotemark{b} (SIS) & $-$3.23 & $-$3.45 & $-$3.74 & $-$4.23 
	& $-$5.04\\
LCDM\tablenotemark{c} ($\alpha=1$) & $-$6.12 & $-$6.17 & $-$6.28 & $-$6.54 
	&$-$7.09\\
LCDM\tablenotemark{c} ($\alpha=1.5$) & $-$4.50 & $-$4.62 & $-$4.80 & $-$5.14 
	& $-$5.77\\
LCDM\tablenotemark{c} ($\alpha=2$) & $-$2.74 & $-$2.90 & $-$3.12 & $-$3.48 
	& $-$4.08\\
LCDM\tablenotemark{c} (SIS) & $-$3.04 & $-$3.24 & $-$3.51 & $-$3.96 
	& $-$4.69\\
LWDM\tablenotemark{d} ($\alpha=1$) & $-$7.45 & $-$7.46 & $-$7.50 & $-$7.59 
	&$-$7.80\\
\enddata

\tablenotetext{a}{SCDM: $\Omega_m = 1$, $\Omega_\Lambda=0$, $\sigma_8 = 0.5$,
	$c_1 (z=0) = 5$ for $\alpha = 1$.}
\tablenotetext{b}{OCDM: $\Omega_m = 0.3$, $\Omega_\Lambda=0$, $\sigma_8 = 
	0.85$, $c_1 (z=0) = 9$ for $\alpha = 1$.}
\tablenotetext{c}{LCDM: $\Omega_m = 0.3$, $\Omega_\Lambda=0.7$, $\sigma_8 = 
	0.95$, $c_1 (z=0) = 7$ for $\alpha = 1$.}
\tablenotetext{d}{LWDM: $\Omega_m = 0.3$, $\Omega_\Lambda=0.7$, $\sigma_8 = 
	0.95$, $c_1 (z=0) = 3.5$ for $\alpha = 1$.}
\tablecomments{The GNFW parameter $\alpha$ is defined in equation 
        (\ref{gnfw}). SIS = singular isothermal sphere. The source object
	is at $z_s = 1.5$. The Hubble constant is $h = 0.7$. The 
        magnification bias is not included, so $P(\Delta\theta) = 
	P(>\Delta\theta)$ is the intrinsic lensing probability.}

\end{deluxetable}

\end{document}